\documentclass[showpacs,preprintnumbers,amsmath,amssymb]{revtex4}

\usepackage{graphicx}
\usepackage{epsfig}

\usepackage{longtable}
\usepackage{longtable}

\begin{document}

\title{Q-alpha values in superheavy nuclei from the deformed
  Woods-Saxon model.}

\author{P.~Jachimowicz}
 \affiliation{Institute of Physics,
University of Zielona G\'{o}ra, Szafrana 4a, 65516 Zielona
G\'{o}ra, Poland}

\author{M.~Kowal} \email{m.kowal@fuw.edu.pl}
\affiliation{National Centre for Nuclear Research, Ho\.za 69,
PL-00-681 Warsaw, Poland}

\author{J.~Skalski}
\affiliation{National Centre for Nuclear Research, Ho\.za 69,
PL-00-681 Warsaw, Poland}

\date{\today}

\begin{abstract}

   Masses of superheavy (SH) nuclei with $Z=98-128$, including
   odd and odd-odd nuclei, are systematically calculated within the
  microscopic-macroscopic model based on the deformed Woods-Saxon potential.
  Ground states are found by minimizing energy over deformations and
  configurations. Pairing in odd particle-number systems is treated
  either by blocking or by adding the BCS energy of the odd quasiparticle.
 Three new parameters are introduced which may be interpreted as the
  constant mean pairing energies for even-odd, odd-even and odd-odd nuclei.
  They are adjusted by a fit to masses of heavy nuclei.
  Other parameters of the model, fixed previously by fitting masses of
  even-even heavy nuclei, are kept unchanged.
  With this adjustment, the masses of SH nuclei are predicted and then
  used to calculate $\alpha$-decay energies to be compared to known measured
  values. It turns out that the agreement between calculated $Q_{\alpha}$
  values with data in SH nuclei is better than in the region of the mass fit.
  The model overestimates $Q_{\alpha}$ for $Z=111-113$.
  Ground state (g.s.) configurations in some SH nuclei hint to a possible
  $\alpha$-decay hindrance. The calculated configuration-preserving
  transition energies show that in some cases this might explain discrepancies,
  but more data is needed to explain the situation.

\end{abstract}


\maketitle

\section{INTRODUCTION}

    Most of the currently known heaviest nuclei, in particular all beyond
  $Z=114$, decay via a sequence of alpha particle emissions
  \cite {O1,GSI,GSI2,GSI3,LBL,O2,Rudolph}.
  Energy release in an $\alpha$-decay of a nucleus with $Z$ protons and
 $N$ neutrons, $Q_{\alpha}(Z,N)$, is directly related to nuclear masses
\begin{equation}
 Q_{\alpha} (Z,N) = M(Z,N)-M(Z-2,N-2)-M(2,2) .
\end{equation}
 Hence energies $E_{\alpha}$ \cite{Ealpha} measured in a chain of
 $\alpha$-decays provide a link between masses of parent and daughter
 nuclei if they can be identified as g.s to g.s transitions. They can
  also determine a newly created nuclide when $E_{\alpha}$ of one of
 the consecutive decays matches the value characteristic of an already known
 parent isotope.
 Besides providing a hint for the identification of new elements, $Q_{\alpha}$
 values are the main factor determining the half-life with respect to the
 $\alpha$-decay.
   Since these half-lives directly relate to the detection pattern,
   a possibly accurate determination of $Q_{\alpha}$ is important for
  the search for new elements.  Finally, although many masses of SH isotopes
  are unknown, this observable provides a test of a local dependence of
 theoretical masses on $Z$ and $N$.

   While the calculations of masses for even-even nuclei are readily available
 in the literature, similar systematic calculations for the odd and odd-odd
 systems are less frequent. Here, we report such calculations for heavy and
 SH nuclei within the microscopic-macroscopic model based on the deformed
 Woods-Saxon potential \cite{WS}. This model was widely applied to many
 problems of nuclear structure over many years.
  Recently, in a version adjusted to heavy nuclei \cite{WSpar},
 we used it to reproduce data on first \cite{Kow}, second \cite{IIbarriers}
 and third barriers \cite{IIIbarriers1,IIIbarriers2}
 and on second minima  \cite{kowskal} in actinides and to predict
 ground states and saddle-points in superheavy nuclei up to $Z=126$
 \cite{archive}.
 The general motivation of our study is to sharpen predictions of the model,
 i.e. masses, $Q_{\alpha}$ values and fission barriers,
 by accounting for sufficiently many deformations (which, for
 technical reasons, was not always practical in the past).
 The results obtained up to now reveal the importance of including
 some deformations, neglected in the previous calculations.
 This concerns especially studies of first, second and third fission barriers.
 In the region of SH nuclei, the predicted abundance of triaxial saddle
 points for $Z\geq 120$ \cite{archive,BROSKAL} calls into question all
 calculations assuming axial symmetry done previously.

 In the present paper we continue along this line by extending our model, which
 up to now was applied mainly to even-even nuclei, to odd and odd-odd nuclei.
 To be sure, the Woods-Saxon model was used for odd SH nuclei previously,
 see for example \cite{LojBar,Cwiok3,ParSob}. However, there are important
  differences between the present study and the previous ones:
  a different version of macroscopic energy giving different results, more
  restricted equilibrium shapes and fewer nuclei were studied in \cite{LojBar,Cwiok3};
  the study of ground and excited states in \cite{ParSob} was
  performed solely without blocking.

 In extending the model we prefer to keep all essential parameters fixed in
 \cite{WSpar} unchanged.
 The extension of the microscopic part consists in calculating the shell and
 pairing correction energy for a system with an odd number of nucleons.
 This is done in two ways, differing by a treatment of the odd particle.
  The macroscopic part is modified by including an additional average pairing
 energy contribution, different for even-odd, odd-even and odd-odd nuclei.
 These contributions are chosen
 as constants and fixed by a fit to the masses of trans-lead nuclei known
  in 2003, in analogy to the fit for even-even nuclei done in \cite{WSpar}.
 After that, the ground states
 of 1364 nuclei, from $Z=98$ to $Z=128$, are determined by energy
 minimization over configurations with zero or one blocked particle over
 axially-symmetric deformations.
 The $\alpha$-decay energies of SH nuclei calculated from these masses are
 compared to the measured values, including recent isotopic chains
  for $Z=117$ \cite{Oga2012}. This allows to appreciate the performance
 of the model outside the region of the original fit and to discuss some
  possible structure effects. We also make comparisons
  with results of some other models.

  A description of our model and calculations is given in section II. The
  results are presented and discussed in section III. Finally, the conclusions
  are summarized in section IV.

\section{The Model}
 Our microscopic-macroscopic model is based on a deformed Woods-Saxon
  potential \cite{WS}.
  In this study we focus on nuclear ground states. Therefore, it is possible
 to confine analysis to axially-symmetric shapes defined by the following
 equation of the nuclear surface:
   \begin{equation}
   \label{shape1}
    R(\theta)= c(\{\beta\})R_0 \left[1+\sum_{\lambda=2}\beta_{\lambda }
   Y_{\lambda 0}(\theta) \right]  ,
  \end{equation}
  where $c(\{\beta\})$ is the volume-fixing factor and $R_0$ is the radius of
 a spherical nucleus. For the macroscopic part we
  used the Yukawa plus exponential model \cite{KN}.
 With the aim of adjusting the model especially for heavy and superheavy nuclei,
 three parameters of the macroscopic energy formula and the pairing strengths
 were determined in \cite{WSpar}
 by a fit to masses of even-even nuclei with $Z\geq84$ and $N>126$ as given in
 \cite{Wapstra}. These parameters were used since then in all our calculations.

 For systems with odd proton or neutron (or both), a standard treatment is
 that of blocking. Considered configurations consist of an odd particle
 occupying one of the levels close to the Fermi level and the rest of the
  particles forming a paired BCS state on the remaining levels.
 The ground state is found by looking for a configuration (blocking
 particles on levels from the 10-th below to 10-th above the Fermi level)
 and deformation giving the energy minimum. In the present study,
 we used this procedure including mass-symmetric deformations
 $\beta_2, \beta_4, \beta_6$ and $\beta_8$, i.e the four-dimensional
 minimization is performed by the gradient method and, for the check,
 on the mesh of deformations:
\begin{eqnarray}
\beta_{2} & = &  \ \  -0.30 \ (0.02) \ 0.32    \nonumber \\
\beta_{4} & = &  \ \  -0.08 \ (0.02) \ 0.18  \nonumber \\
\beta_{6} & = &  \ \  -0.10 \ (0.02) \ 0.12    \nonumber\\
\beta_{8} & = &  \ \  -0.10 \ (0.02) \ 0.12    \nonumber\\
\end{eqnarray}
 Both sets of results are consistent; lower energies from the gradient
 method are treated as final.
 The used deformation set should provide for a fair approximation,
 except for the region of light isotopes
 of elements between Rn and light actinides, which
 show octupole deformation in their ground states.
 The values of parameters from \cite{WSpar} were left unchanged for even-even
  nuclei. For the rest, we introduced three
 new parameters - additive constants which may be interpreted as
 corrections for the mean pairing energy in even-odd, odd-even and
 odd-odd nuclei. These parameters were fixed by a fit to the masses
 of odd-even, even-odd and odd-odd $Z\geq82$ and $N>126$ nuclei taken
 from \cite{Wapstra2003}.

 It is known that the blocking procedure often causes an excessive reduction
  of the pairing gap in systems with odd particle number. One device to avoid
  an excessive even-odd staggering in nuclear binding was to assume
  stronger (typically by $\sim$ 5\%) pairing interaction for
  odd-particle-number systems, see \cite{Gor0}. Since the main predictions
  of this work are $Q_{\alpha}$ values in which the effect of stronger pairing
  in parent and daughter nuclei partially cancels out, we postpone
  for the future a more elaborate treatment of this effect. Instead,
 we performed another calculations of nuclear masses without blocking.
  Shell (and pairing) correction energy of a configuration with an odd neutron
 (or proton) was taken as a sum of the quasiparticle energy of a singly
 occupied level $\sqrt{(\epsilon-\lambda)^2+\Delta^2}$ and the shell
 (and pairing) correction calculated without blocking.
 The latter quantity, as well as the pairing gap $\Delta$ and the Fermi energy
 $\lambda$, are calculated for the odd number of particles, but with the double
  occupation of all levels. This prescription was used before in \cite{ParSob}.
  It gives results similar to those obtained when calculating $\Delta$,
 $\lambda$ and the shell (and pairing) correction for the even system with
 one particle less.
  The calculation without blocking is much simpler and we were
 able to perform a seven-dimensional minimization over axially-symmetric
 deformations
 $\beta_2, \beta_3, \beta_4, \beta_5, \beta_6, \beta_7$ and $\beta_8$.
 Therefore, these results should be reliable also for light actinides.
 As we preferred to avoid a new fit of the macroscopic model parameters,
 also for this model we introduced three additive constants (energy shifts)
 for even-odd,  odd-even and odd-odd nuclei which
  minimize the rms deviation in each of the groups of nuclei.

 \subsection{Odd-odd nuclei}

 Structure of odd-odd nuclei is more complicated than that of odd-$A$
 systems. If we disregard collective vibrations, the ground state
 configuration is a result of coupling the unpaired neutron and proton to
  a total angular momentum. The energy ordering of coupled configurations
 is usually attributed to a residual neutron-proton interaction $V_{np}$.
 In spherical nuclei it is summarized by the empirical Nordheim rule
\cite{Nord}.

 In deformed, axially symmetric nuclei,
 in which the projection of the single-particle angular momentum on the
 symmetry axis $\Omega$ is a good quantum number, the n-p coupling can give
  two configurations with $K=\mid \Omega_p\pm \Omega_n\mid$.
  According to the empirical Gallagher-Moszkowski rule \cite{GM}, the one
  energetically favoured is the spin triplet state.
 The spin structure of both n and p single particle orbitals shows which
  $K$ configuration will be the lower one.
 A collective rotational band is built on each of two bandheads. Energies of
 the band members with angular momentum $I$ are usually presented as
 \cite{Jain}
 \begin{equation}
   E(I,K) = E(n,p) + \frac{\hbar^2}{2{\cal J}} \left[I(I+1)-K^2\right] +
   E_K + (-)^I \delta_{K,0} (E_0+E_a) ,
 \end{equation}
  where $E(n,p)$ represents the mean-field energy of a bandhead configuration,
  the second term is the rotational energy, $E_K$ is the diagonal matrix
  element of $V_{np}$, the last term, combined
  of the Newby shift $E_0$ and the diagonal Coriolis term $E_a$ for
  $\Omega_n=\Omega_p=1/2$, occurs only for $K=0$ bands and
  splits them into two subbands. In such a formula, all
  off-diagonal matrix elements of the interaction $V_{np}$ and of the
  rotor-plus-two-particles Hamiltonian are neglected. From the experimental
   data in rare earth and actinide regions, the Newby shifts $E_0$ and
  Gallagher-Moszkowski shifts, defined as $\Delta E_{GM}=E_{K<}-E_{K>}=
  \hbar^2\Omega_{<}/{\cal J}+E_{b.head}(K_{<})-E_{b.head}(K_{>})$,
   were extracted \cite{Jain,Sood}.
 The former are usually less than 50 keV, while the latter amount to
  100-300 keV.

  The above information is incorporated in mass formulas by defining
  some average (i.e. configuration-independent) neutron-proton energies
   for odd-odd nuclei.
   Their role is to account for the shift in the g.s. energy with respect
   to the value caluclated with blocking or quasiparticle method
   that would simulate on average the terms beyond $E(n,p)$ in Eq. (4).
  For example, in \cite{mol95}, the additional binding $\delta_{np}$ is
  included which amounts to $\sim 200$ keV for odd-odd actinides.
  Although this term is $A$-dependent in \cite{mol95}, one can see that the
  difference in it between actinides and superheavy nuclei is around 20 keV.
  Therefore, as we confine here our model to heavy and superheavy nuclei,
  we assume constant average neutron-proton and average pairing energies.
  This leaves three constants: $h_{oe}$, $h_{eo}$ and $h_{oo}$ (see Table 1 and
  2) that can be fit to odd-$A$ and odd-odd nuclei;
  they correspond to the parameters ${\bar \Delta}_n$, ${\bar \Delta}_p$ and
   ${\bar \Delta}_n+{\bar \Delta}_p-\delta_{np}$ of the model used in
  \cite{mol95}. Thus, we calculate the mass of an odd-odd nucleus within the
  blocking method by adding 1.703 MeV to the micro-macro energy of the optimal
  configuration. This corresponds to the neutron-proton energy of
  $h_{oe}+h_{eo}-h_{oo}=134$ keV (Table 1).

  Since the g.s. configurations must be energetically favored,
  the parent and daughter energy shifts $E_K$ will cancel in large part in
  $Q_{\alpha}$ values.

 \subsection{Configuration hindrance of $\alpha$ transitions}

 Considering a comparison of measured and calculated $\alpha$-decay energies,
  it is important to observe the hindrance of $\alpha$ transitions
  between different configurations in odd-$A$ and odd-odd nuclei. Although
  a degree of this hindrance is surely configuration-dependent,
  if strong enough, it can hide the true $Q_{\alpha}$ value when
  only a few transitions are detected in experiment.
  At present, this is the situation in many heaviest nuclei.

  One can consult the known data to see the magnitude of hindrance.
  For example, the isotopes $^{251}$Fm, $^{253}$No and $^{255}$Rf decay
  primarily to the $9/2^-$ parent g.s. configurations in daughters, which lie
  at the excitation energy of 200-400 keV, with probabilities, respectively,
 87\% \cite{Ahmad}, 96\% \cite{Hes1} and $>90$\% \cite{Hes2}. In 82\% of cases,
 $^{249}$Cf decays to the $9/2^-$ parent configuration;
  the $7/2^+$ g.s. in $^{245}$Cm daughter is populated only in 2.5\% of cases
  \cite{249Cf}. In the decay of $^{251}$Cf, the hindrance of the g.s. to g.s
  transition ($1/2^+\rightarrow 9/2^-$) results in the g.s. band in daughter
  receiving $\sim 15$\% of cases; 2.6\% of those decays goes to the g.s.
  \cite{251Cf}. Much reduced $K$-hindrance is seen in the decay of
  $^{249}$Bk ($7/2^+$): more than 90\% of decays goes to the g.s. rotational
  band in $^{245}$Am, built on the $5/2^+$ configuration, in that
  6.6\% to the g.s \cite{249Bk}.

  Motivated by these examples, to set an upper limit for an underestimate of
 $Q_{\alpha}$, we also calculate apparent $Q_{\alpha}$ values taking
  the parent g.s. configuration as the final state in daughter.
   Such a value is smaller than the true $Q_{\alpha}$ by the
   excitation of the parent g.s. configuration in the daughter.
   Gallagher shifts also mostly cancel in such transition energies.


 \section{Results and discussion}
 The quality of the mass fit is summarized in Tables I and II where
 the deviations from
 the experimental masses are given for each of the four groups of nuclei.
 The data are taken from \cite{Wapstra2003}, also for even-even nuclei.
 Statistical parameters of the fit for this group are different
  in Tables I and II because of different deformations included. Deviations
  in Table II slightly differ from those in \cite{WSpar} because
 we used here a larger number of data from \cite{Wapstra2003}.
 One can observe that the model with blocking is worse for even-odd and
 odd-even systems than for the even-even ones; the quality deteriorates
 further for odd-odd systems.
 A different situation occurs for the results without blocking:
 the worst case are the odd-even systems; odd-odd masses are rather well
 described. The differences in $\delta_{RMS}$ between groups of nuclei
 may show a need to refit some of the parameters fixed for even-even nuclei,
 but this requires more study.

 One can observe that our local fit is better that those resulting from
 the self-consistent models. For example, after
adding to the typical Skyrme forces the phenomenological Wigner
term, microscopic contact pairing force and correction for
spurious collective energy, the root-mean square deviation equal
0.58 MeV has been obtained in a global calculation, see ref
\cite{Gor1,Gor2}. In another Hartree-Fock-Bogoliubov (HFB) model \cite{Gor3},
aimed at fitting simultaneously masses and fission data,
a phenomenological correction for collective vibrations allowed to obtain
the r.m.s deviation of 0.729 MeV. Recently HFB calculations
via refitting to the 2012 Atomic Mass Evaluation (AME) and varying the symmetry coefficients
gave in the best case a value of 0.54 MeV r.m.s \cite{Gor13}.

  Macroscopic-microscopic global calculations of
 nuclear masses made by P. Moller and co-workers
\cite{mol95} give the r.m.s error 0.669 MeV for nuclei ranging
from Oxygen to Hassium and  0.448 MeV in the case of nuclei above
$N=65$. A phenomenological formula with the 10 free parameters
by Duflo and Zuker  \cite{Duflo1,Duflo2} gives mass estimates with
the 0.574 MeV r.m.s error. Recently, authors of \cite{Liu} achieved
 the r.m.s deviation of 0.34 keV in a fit to masses of 2149 nuclei,
 however, at the cost of including many corrections with often a rather
  obscure physical meaning.

\begin{table}
\begin{ruledtabular}
\caption{ Statistical parameters of the fit to masses in the model
with
 blocking in separate groups of even-even, odd-even, even-odd and odd-odd
 heavy nuclei:
  the number $N$ of nuclei in the group, the energy shift $h$,
 the average discrepancy $<\mid M^{exp}-M^{th}\mid>$,
 the maximal difference $max \mid M^{exp}-M^{th}_{f}\mid$, the $rms.$
 deviation $\delta_{\rm RMS}$. Experimental data taken from \cite{Wapstra2003}. All
quantities are in MeV, except for the number of nuclei N. }
\label{1}
\begin{tabular}{cccccc}

  & {e - e}& {o - e} & {e - o}& {o - o} \\
\cline{2-5}

\noalign{\smallskip} \noalign{\smallskip}

       N                                        & 74     & 56    &    69  & 53      \\
 \hline
 $h$                                            &  0.0   & 1.013 &
0.824 & 1.703   \\
 $<\mid M^{th}- M^{exp}\mid>$                   & 0.212  & 0.340 &
0.356 & 0.566   \\
 $Max \mid M^{th}- M^{exp}\mid$                 & 0.833  & 0.836 &
1.124 & 1.387   \\
$\delta_{\rm RMS}$                              & 0.284  & 0.425 &
0.435 & 0.666   \\

\noalign{\smallskip} \noalign{\smallskip}
\end{tabular}
\end{ruledtabular}
\end{table}
\begin{table}
\begin{ruledtabular}
\caption{The same as in Table \ref{1} but for the method without
blocking.} \label{2}
\begin{tabular}{cccccc}

 & {e - e}& {o - e} & {e - o}& {o - o} \\
\cline{2-5}
\noalign{\smallskip} \noalign{\smallskip}

       N                                        & 74    & 56     &    69
 & 53     \\
  \hline
 $h$                                            & 0.0  & -0.751 &
0.268 & 0.234    \\
 $<\mid M^{th}- M^{exp}\mid>$                   & 0.187 & 0.460  &
0.273 & 0.295  \\
 $Max \mid M^{th}- M^{exp}\mid$                 & 0.652 & 1.398  &
0.892 & 0.853  \\
$\delta_{\rm RMS}$                              & 0.251 & 0.551  &
0.343 & 0.366 \\

\noalign{\smallskip} \noalign{\smallskip}
\end{tabular}
\end{ruledtabular}
\end{table}
\begin{figure}
\includegraphics[width=\columnwidth]{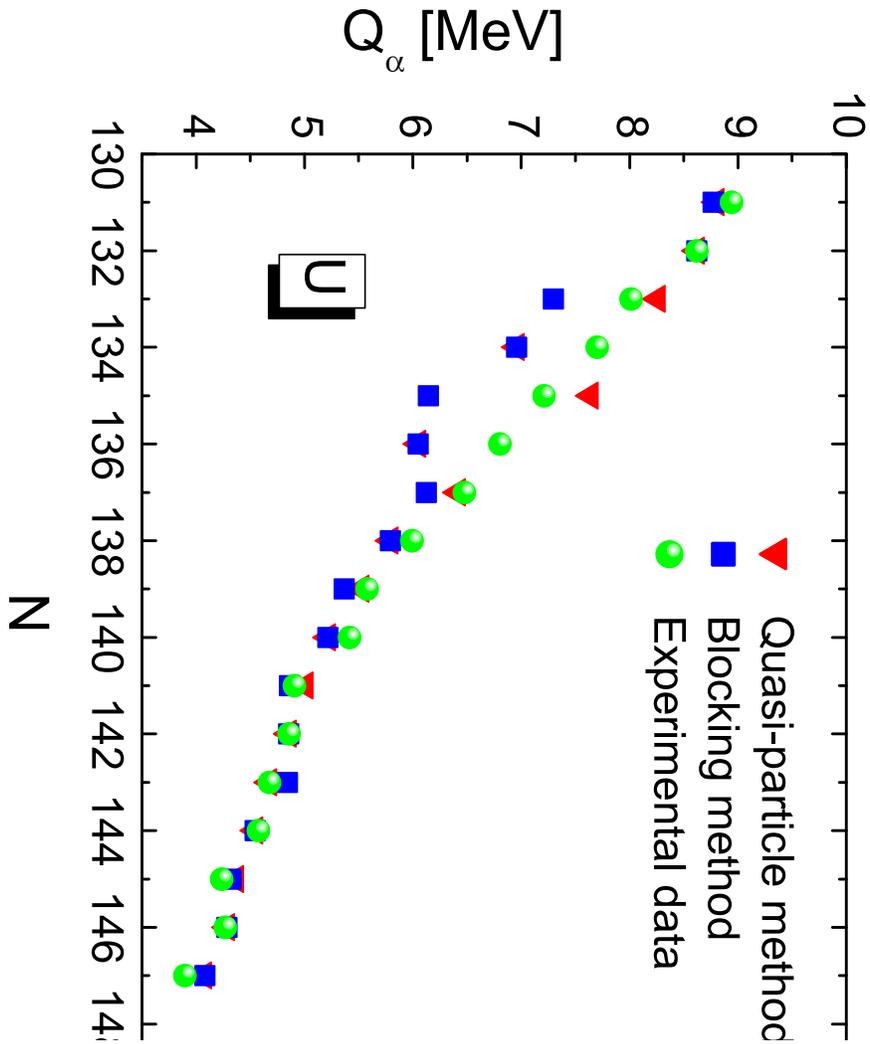}
\caption{\label{fig1}
 $Q_{\alpha}$ values for U ($Z=92$) nuclei:
 circles - experiment, squares - model with blocking, triangles - model
 without blocking. }
\end{figure}
\begin{figure}
\includegraphics[width=\columnwidth]{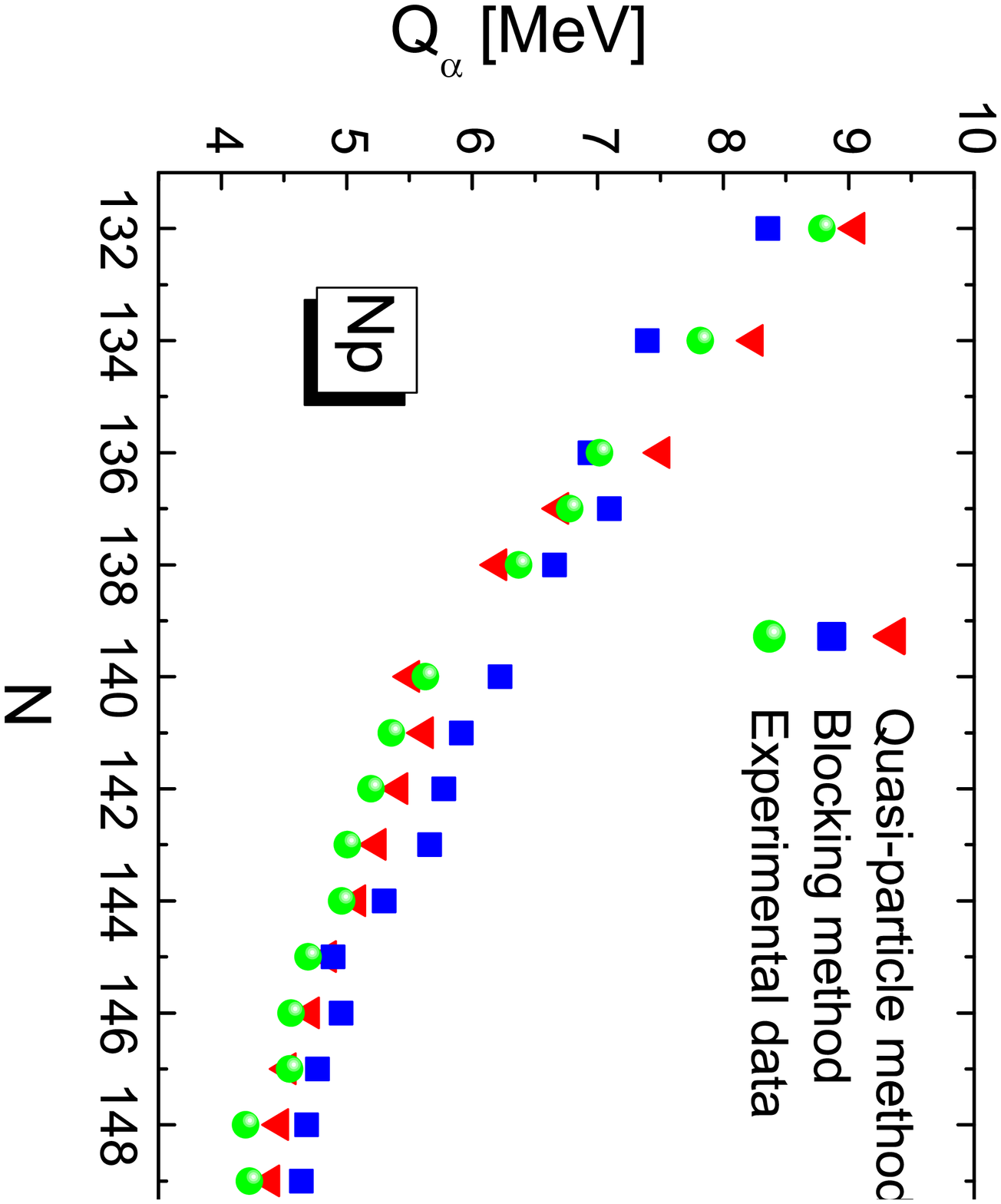}
\caption{\label{fig2} As in Fig. 1, but for Np ($Z=93$) isotopes.}
\end{figure}
  The calculated and measured $Q_{\alpha}$ values for U and Np isotopes are
 shown in Fig. 1 and 2. They illustrate the quality of the model
 for nuclei from the region of the fit. We did not choose the best cases;
 on the contrary, the values calculated with blocking for the Np isotopes
 are systematically overestimated.
 Calculations without blocking are clearly better for this isotopic chain.
 In uranium nuclei, both calculations agree well with the data for $N>136$.
 For smaller $N$, there are quite large discrepancies. In the case of the
 model with blocking (but not the other one) these may come from discarding
 the reflection asymmetry.

 To check the effect of the octupole deformation
 we have chosen the nucleus $^{225}$U (see Fig. \ref{fig1}).
 We added deformations $\beta_{30},\beta_{50},\beta_{70}$ to
the original grid and conducted the seven-dimensional minimization
 using the method with blocking. The reflection-asymmetric deformations of the
 found minimum were $\beta_{30}=0.122 ,\beta_{50}=0.045 ,\beta_{70}=0.010$
 and $\beta_{30}=0.089 ,\beta_{50}=0.043 ,\beta_{70}=0.030 $ in the
 parent and daughter nuclei, respectively. They reduce
 the g.s. energy by 0.6 MeV in the parent nucleus
 and by about 1.5 MeV in the daughter. This leads
 to the increase in alpha-decay energy by about 0.8 MeV,
 giving $Q_{\alpha}=$8.2 MeV which agrees very well with
 the measured 8.00 MeV. Final results for $^{225}$U (together with
the corresponding results of the quasiparticle method) are
summarized in Tables III and IV. It is likely that
  the results with blocking would improve also for some other neutron
deficient nuclei.

  For a set of 204 nuclei from the region of the fit one can compare
  the calculated and experimental $Q_{\alpha}$ values. The average
  deviation $<\mid Q_{\alpha}^{exp}-Q_{\alpha}^{th}\mid>$ amounts to
  326 keV for the calculation with blocking and 225 keV for the
  quasiparticle method; the rms deviations are, respectively, 426 and
  305 keV. Thus, the quasiparticle method gives a better agreement
  with experimental $Q_{\alpha}$ values in the region of the fit.

\begin{table}
\begin{ruledtabular}
\caption{Effect of octupole deformations within blocking method.}
\label{3}
\begin{tabular}{cccccccccc}
 & $M^{th}$& {$\beta_{20}$} & $\beta_{30}$ & $\beta_{40}$ &
$\beta_{50}$ & $\beta_{60}$ & $\beta_{70}$ &$\beta_{80}$ \\

 \noalign{\smallskip} \noalign{\smallskip}
  \hline
$^{225}U$  & 25.490 & 0.134 & 0.123         & 0.074  & 0.045 &
0.012 &
0.011  & -0.004  \\
$^{225}U$  & 26.081 & 0.157 &       -       & 0.107  & -     & 0.044 &
  -       & -0.009 \\
$^{221}Th$ & 14.878 & 0.110 & 0.092         & 0.073  & 0.045 & 0.022 &
0.028  & -0.001  \\
$^{221}Th$ & 16.387 & 0.110 &       -       & 0.083  & -     & 0.039 &
  -       & -0.022 \\

\noalign{\smallskip} \noalign{\smallskip}
\end{tabular}
\end{ruledtabular}
\end{table}

\begin{table}
\begin{ruledtabular}
\caption{Effect of octupole deformations within quasi-particle
method.} \label{4}
\begin{tabular}{cccccccccc}
 & $M^{th}$& {$\beta_{20}$} & $\beta_{30}$ & $\beta_{40}$ &
$\beta_{50}$ & $\beta_{60}$ & $\beta_{70}$ &$\beta_{80}$ \\

 \noalign{\smallskip} \noalign{\smallskip}
 \hline
$^{225}U$  & 26.463&  0.131& 0.114 &  0.074 & 0.044 & 0.014& 0.015
&
-0.003 \\
$^{225}U$  & 27.278&  0.151&       -            & 0.103 & -      & 0.039 &
      -       & -0.015 \\
$^{221}Th$ & 15.778&  0.103& 0.093 &  0.069 & 0.047 & 0.021& 0.030 &
-0.002 \\
$^{221}Th$ & 17.247&  0.104&       -            & 0.077 & -      &  0.038
&       -       & -0.012 \\

\noalign{\smallskip} \noalign{\smallskip}
\end{tabular}
\end{ruledtabular}
\end{table}
 Starting with Fig. 3, we present the predictive part of the model:
 most of the masses of SH nuclei involved in $\alpha$-decay energies
  were {\it not included} in the fit.
 The $Q_{\alpha}$ values calculated with blocking are compared to
  experimental data in Table V. We used mostly the data from
 \cite{masstable2013}, but in a few cases relied on other sources.
 In particular, the $Q_{\alpha}$ values in chains $^{293,294}$117 were based
 on \cite{O2,Oga2012} and deduced from the upper range of energies
  $E_{\alpha}$ when such a range exceeded the energy resolution of the
  detector (see Table II in \cite{O2}). The $Q_{\alpha}$ values are shown
 for a wider range of $N$ in Fig. 3, separately for even- and odd-Z nuclei.
  In Fig. 4-7 the values calculated with and without blocking are
  shown vs experimental data for $Z=103,107,108,113$ nuclei.
 One should bear in mind  that calculated decay energies {\it are independent}
 of the fitted energy shifts (average pairing energies), denoted $h$
  in Tables I, II.

 The calculated $Q_{\alpha}$ vs $N$ plots (Fig. 3) show a pronounced rise
 for $N$ overstepping 184 and smaller ones at $N=152$ and 162.
 They signal particularly well bound systems at these neutron numbers.
  The first one is connected with the magic spherical configuration
 (not yet tested by experiment) and the other two with the particularly
 stable prolate deformed configurations, corresponding to prominent gaps
  in the s.p. spectrum - see Fig. \ref{fig10}.
  One can notice that the maximum around $N=162$ becomes wider for
  larger $Z$ and some other maxima appear between $N=160$ and 184,
 especially for $Z\geq 120$. On average, $Q_{\alpha}$ values increase with
 $Z$ at constant $N$. A larger than average increase is predicted above $Z=108$
  for $N\leq170$ and is related to the deformed proton subshell -
  see Fig. \ref{fig10}.
  It is visible in Fig. 3 as a larger gap between the plots
  for fixed $Z$, especially between $Z=108$ and $Z=110$.
  A number of smaller proton shell effects is predicted for limited
  ranges of $N$, like for $Z=114$ around $N=180$.

\begin{figure}
\includegraphics[width=\columnwidth]{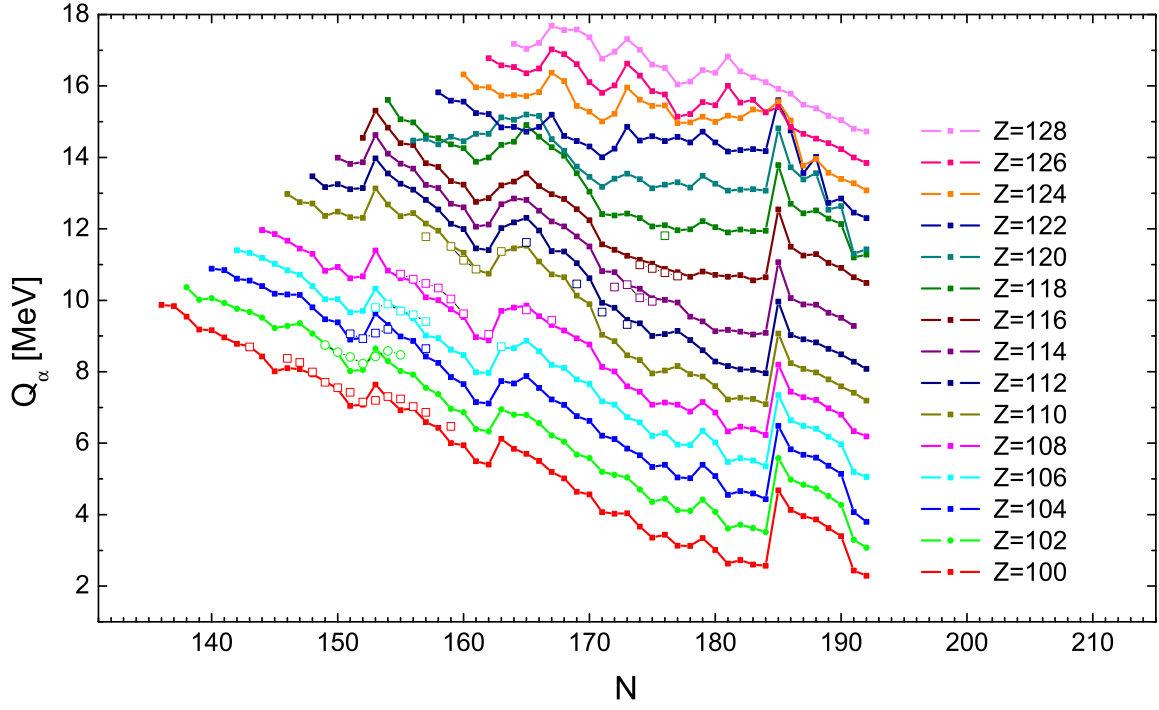}
\includegraphics[width=\columnwidth]{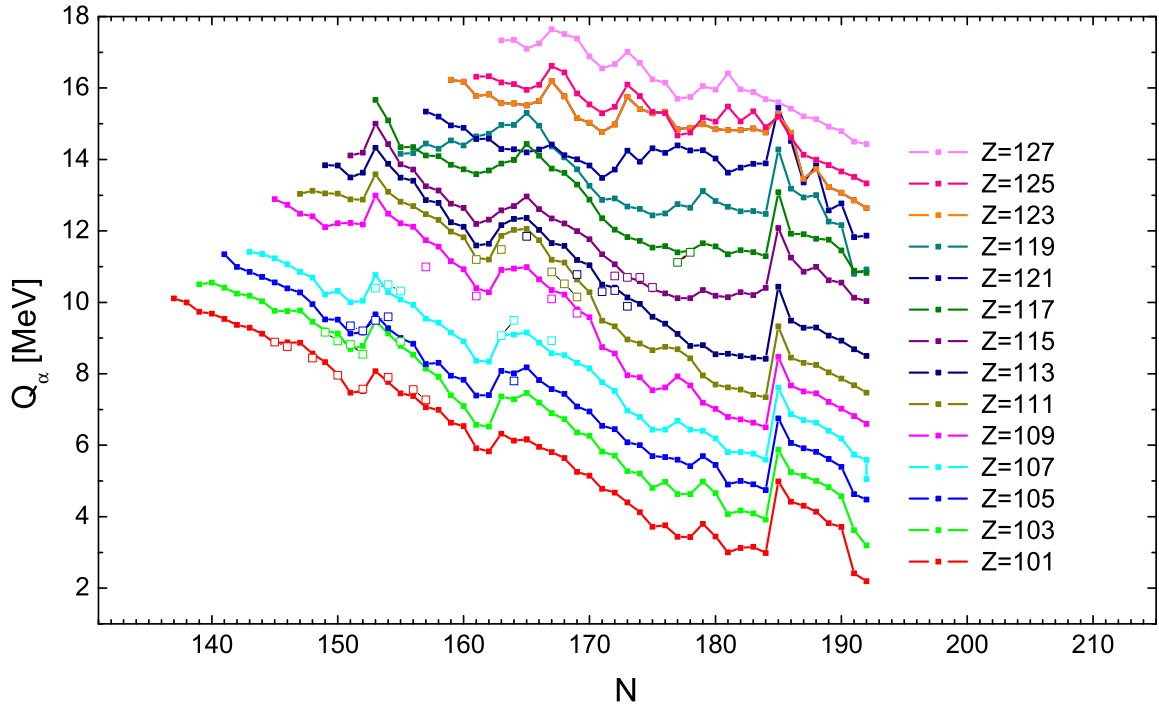}
\caption{\label{fig3}
 $Q_{\alpha}$ values calculated with blocking vs experimental data.
 Explicit values, including the source for the experimental
 data, are given in Table V. }
\end{figure}

   The rise in $Q_{\alpha}$ when going through $N=152$ is supported by the
   data for $Z\geq 100$, but is much more gentle than the one calculated
   with blocking. A jump in $Q_{\alpha}$ across the closed "subshell" is much
   reduced in calculations without blocking, see Fig. 4-7.
   In the data, transition energy increases for at least
   two successive neutron numbers (this means for 153 and 154).
  The increase in $Q_{\alpha}$ above $N=162$ is best seen in the data
   for Hs and Ds in Fig. 3, and roughly consistent with calculations;
   it is smaller than predicted for Rg.
   The proton shell effect at $Z=108$ is seen in the data, but slightly
   smaller than calculated.

%
%

\begingroup

\squeezetable


\begin{table*}
\label{5}

\caption{ Results of the calculations with blocking. In successive
 columns are given: proton number $Z$,
 neutron number $N$, mass number $A$, parent quadrupole deformation
 $\beta_{20}$, experimental value $Q_{\alpha}^{exp}$ from \cite{masstable2013},
  \cite{O2,Oga2012} (O), \cite{Rudolph} (R) and \cite{Gupta2005} (*),
 calculated g.s. to g.s. values $Q_{\alpha}(gs\rightarrow gs)$,
 the parent g.s. configuration $\pi \{\Omega\}_{P}(gs)$ specified by the parity
  and $\Omega$-quantum numbers (multiplied by 2, P - protons, N - neutrons),
 the daughter g.s. configuration $\pi \{\Omega\}_{D}(gs)$ and the calculated
 decay energy $Q_{\alpha}$($\pi \{\Omega\}_{P}=\pi \{\Omega\}_{D}$) for the
 configuration-preserving transition.
    }

 \begin{ruledtabular}

\begin{tabular}{|c|c|c|c|c|c|cc|cc|c|}

 Z & N & A & $\beta_{20}$ & $Q_{\alpha}^{exp}$ & $Q_{\alpha}(gs \rightarrow gs)$ &
 $\pi \{\Omega\}_{P}(gs)$ & & $\pi \{\Omega\}_{D}(gs)$ & & $Q_{\alpha}$($\pi \{\Omega\}_{P}=\pi
  \{\Omega\}_{D}$)   \\
 & & & & & & P  & N  &    P & N  &  \\

\noalign{\smallskip}\hline\noalign{\smallskip}


   119 &   178 & 297 & -0.10 &   & 12.64   &   -1  &     &   -3  &     &   12.57   \\
\hline
   118 &   176 & 294 & -0.09 & 11.81  & 12.09   &    &    &     &     &           \\
\hline
   117 &   177 & 294 & -0.09 & 11.12$^O$ & 11.32   &   -3  &   1   &   -5  &   1   &   11.25   \\
   117 &   176 & 293 & -0.09 & 11.36$^O$ & 11.53   &   -3  &     &   -5  &     &   11.45   \\
\hline
   116 &   177 & 293 & -0.08 & 10.68  & 10.80   &     &   1   &     &   1   &   10.80   \\
   116 &   176 & 292 & -0.08 & 10.77  & 10.92   &     &      &      &      &           \\
   116 &   175 & 291 &  0.08 & 10.89  & 11.01   &     &   1   &     &   5   &   10.89   \\
   116 &   174 & 290 & -0.09 & 10.99  & 11.14   &     &    &     &      &           \\
\hline
   115 &   175 & 290 & -0.08 & 10.42$^O$  & 10.41   &   -5  &   1   &   -7  &   5   &   10.11   \\
   115 &   174 & 289 & -0.09 & 10.69$^O$  & 10.60   &   -5  &      &   -3  &     &   10.56   \\
   115 &   173 & 288 &  0.08 & 10.70$^R$  & 10.76   &   -1  &   5   &   -3  &   5   &   10.54   \\
   115 &   172 & 287 & -0.11 & 10.74  & 11.01   &   -5  &      &   -3  &     &   10.61   \\
\hline
   114 &   175 & 289 &  0.09 &  9.97 & 10.00   &      &   1   &      &   -15 &    9.93   \\
   114 &   174 & 288 &  0.09 & 10.07  & 10.32   &       &       &       &       &           \\
   114 &   173 & 287 &  0.09 & 10.16  & 10.44   &      &   5   &      &   5   &   10.44   \\
   114 &   172 & 286 & -0.12 & 10.37  & 10.80   &      &      &      &      &           \\
\hline
   113 &   173 & 286 &  0.09 &  9.89$^O$  & 10.13   &   -7  &   5   &   -1  &   5   &   9.68    \\
   113 &   172 & 285 &  0.14 & 10.33$^O$  & 10.45   &   -3  &     &   11  &      &   10.27   \\
   113 &   171 & 284 &  0.14 & 10.30$^R$  & 10.52   &   -3  &   5   &   -9  &   5   &   10.34   \\
   113 &   170 & 283 &  0.15 &        & 11.09   &   -3  &      &   -3  &      &   11.09   \\
   113 &   169 & 282 &  0.21 & 10.78  & 11.22   &   -1  &   1  &   -3  &   5  &   10.71   \\
   113 &   168 & 281 &  0.21 &         & 11.56   &   -1  &      &   -3  &      &   11.34   \\
   113 &   167 & 280 &  0.21 &         & 11.60   &   -1  &   5  &   -3  &   3  &   11.25   \\
   113 &   166 & 279 &  0.21 &         & 12.02   &   -1  &      &   -3  &      &   11.81   \\
   113 &   165 & 278 &  0.21 & 11.85  & 12.33   &   -1  &   3   &   -3  &   -13 &   11.56   \\
\hline
   112 &   173 & 285 &  0.11 & 9.32  &  9.48    &    &  -15   &     &   5   &   9.35    \\
   112 &   172 & 284 &  0.13 &        &  9.77    &    &        &     &       &           \\
   112 &   171 & 283 &  0.13 & 9.67*  & 9.91   &    &   5   &     &   9   &    9.91   \\
   112 &   170 & 282 &  0.14 &         & 10.69  &    &       &     &       &      \\
   112 &   169 & 281 &  0.20 & 10.46  & 11.07   &     &   1   &     &   5   &   10.78   \\
   112 &   168 & 280 &  0.19 &         & 11.38   &    &       &     &       &           \\
   112 &   167 & 279 &  0.20 &         & 11.45   &    &   5   &     &   3   &   11.31   \\
   112 &   166 & 278 &  0.20 &         & 11.86   &    &       &     &       &           \\
   112 &   165 & 277 &  0.21 & 11.62  & 12.21   &    &   3   &
    & -13   &   11.66   \\
\hline
   111 &   171 & 282 &  0.14 &   &  9.49   &   -1  &   5   &   11  &   9   &    9.24   \\
   111 &   170 & 281 &  0.15 &   & 10.36   &  11  &      &   11  &      &   10.36   \\
   111 &   169 & 280 &  0.16 & 10.15$^R$  & 10.77   &  -9   &   5   &   11   &   5   &   10.03   \\
   111 &   168 & 279 &  0.20 & 10.52  & 11.13   &  -3   &   &   11  &     &   10.57   \\
   111 &   167 & 278 &  0.21 & 10.85  & 11.18   &  -3   &   5   &   11  &   3   &   10.51   \\
   111 &   166 & 277 &  0.21 &         & 11.64   &  -3   &       &   11  &       &   11.11   \\
   111 &   165 & 276 &  0.21 &         & 11.98   &  -3   &   3   &   11  & -13   &   10.95   \\
   111 &   164 & 275 &  0.22 &         & 12.06   &  -3   &       &   11  &       &   11.56   \\
   111 &   163 & 274 &  0.22 & 11.48  & 11.91   &  -3   &  -13  &   11  &  9  &   10.60   \\
   111 &   162 & 273 &  0.23 &         & 11.23   &  -3   &       &   11  &     &   10.73   \\
   111 &   161 & 272 &  0.23 & 11.20  & 11.28   &  -3   &   7   &   11  &   9   &   10.76   \\
\hline
   110 &   169 & 279 &  0.18 &   &  10.19   &     &   9   &     &   5   &   10.03   \\
   110 &   163 & 273 &  0.22 & 11.37  & 11.49   &     &  -13  &     &   9   &   10.52   \\
   110 &   162 & 272 &  0.23 &         & 10.75   &     &       &     &       &           \\
   110 &   161 & 271 &  0.23 & 10.87  & 10.80   &     &   9   &     &   7   &   10.77   \\
   110 &   160 & 270 &  0.23 & 11.12  & 11.38   &     &       &       &       &           \\
   110 &   159 & 269 &  0.23 & 11.51  & 11.61   &     &   9   &     &   -11 &   11.44   \\
   110 &   158 & 268 &  0.23 &         & 11.94   &     &       &     &       &           \\
   110 &   157 & 267 &  0.24 & 11.78  & 12.11   &     &   3   &     &   1   &   12.10   \\
\hline
   109 &   169 & 278 &  0.19 &   9.69$^O$ &  9.78   &   11  &   9   &   -1  &   5 &   9.27    \\
   109 &   167 & 276 &  0.21 & 10.10$^R$  & 10.17   &   11  &   5   &   -5  &   3   &   9.46    \\
   109 &   166 & 275 &  0.21 &        & 10.67   &   11  &     &  -5   &     &   9.79    \\
   109 &   165 & 274 &  0.21 &        & 11.01   &   11  &   3 &  -5   &  -13  &   9.63    \\
   109 &   164 & 273 &  0.22 &   & 11.11    &  11   &      &  -5  &      &      9.98    \\
   109 &   163 & 272 &  0.22 &   & 11.02    &  11   & -13  &  -5  &  9   &      9.03    \\
   109 &   162 & 271 &  0.23 &   & 10.27    &  11   &      &  -5  &      &      9.00    \\
   109 &   161 & 270 &  0.23 & 10.18  & 10.33    &  11   &   9   &  -5  & 7  &      9.01    \\
   109 &   160 & 269 &  0.23 &   & 10.95    &  11   &      &  -5  &      &      9.58    \\
   109 &   159 & 268 &  0.23 &   & 11.18   &   11  &  9  &  -5   &  -11  &    9.62   \\
   109 &   158 & 267 &  0.23 &   & 11.56   &   11  &     &  -5   &       &   10.17   \\
   109 &   157 & 266 &  0.24 & 11.00  & 11.73   &   11  &   3   &  -5   &   1   &  10.32    \\
\hline
   108 &   167 & 275 &  0.21 & 9.44  &  9.26    &    &   5   &    &   3   &   9.12    \\
   108 &   166 & 274 &  0.22 &   &  9.55    &    &       &    &       &           \\
   108 &   165 & 273 &  0.22 & 9.73  &  9.89    &    &   3   &     &  -13  &   9.44    \\
   108 &   164 & 272 &  0.23 &   &  9.80    &    &       &     &       &           \\
   108 &   163 & 271 &  0.23 &   &  9.72    &    &   -13 &     &   9   &   8.72    \\
   108 &   162 & 270 &  0.23 & 9.05  &  8.87    &       &       &       &       &           \\
   108 &   161 & 269 &  0.24 &   &  8.91    &     &   9   &     &   7   &   8.82    \\
   108 &   160 & 268 &  0.24 &  9.62  &  9.51    &       &       &       &       &           \\
   108 &   159 & 267 &  0.24 & 10.04  &  9.75    &    &   7   &     &  -11  &    9.63   \\
   108 &   158 & 266 &  0.24 & 10.35  &  10.04    &       &       &       &       &           \\
   108 &   157 & 265 &  0.25 & 10.47  &  10.17    &   &  -11  &    &   3   &   10.15   \\
   108 &   156 & 264 &  0.24 & 10.59  &  10.58    &       &       &       &       &           \\
   108 &   155 & 263 &  0.25 & 10.73  &  10.67    &    &   1   &    &   1 &  10.67    \\
\hline
   107 &   167 & 274 &  0.21 & 8.93  &  8.61    &   -1  &   5   &  -5   &   3   &   8.44    \\
   107 &   166 & 273 &  0.22 &   &  8.83    &   -5  &       &  9    &       &   8.79    \\
   107 &   165 & 272 &  0.22 &   &  9.16    &   -5  &   3   &   9   &  -13  &   8.71    \\
   107 &   164 & 271 &  0.23 & 9.49  & 9.10   &   -5  &      &   9   &      &   8.96    \\
   107 &   163 & 270 &  0.23 & 9.06  & 9.09   &   -5  &  -13  &   9   &   9   &   7.92    \\
   107 &   162 & 269 &  0.23 &   &  8.26  &   -5  &       &   9   &       &   7.93    \\
   107 &   161 & 268 &  0.24 &   &  8.33    &   -5  &    9  &   9   &   7   &   7.87    \\
   107 &   160 & 267 &  0.24 &   &  8.92    &   -5  &      &   9   &    &   8.45    \\
   107 &   159 & 266 &  0.24 &   &  9.15    &   -5  &   7   &   9   &  -11  &   8.56    \\
   107 &   158 & 265 &  0.24 &   &  9.45    &  -5   &      &   9   &    &   8.93    \\
   107 &   157 & 264 &  0.25 &   &  9.57    &  -5   &   -11 &   9   &   3   &   9.00    \\
   107 &   156 & 263 &  0.25 &   &  9.99    &  -5   &       &   9   &       &     9.44  \\
   107 &   155 & 262 &  0.25 & 10.32  & 10.11    &   -5  &   1   &   9   &   1   &   9.54    \\
   107 &   154 & 261 &  0.24 & 10.50  & 10.43   &  -5   &      &   9   &     &   9.89    \\
   107 &   153 & 260 &  0.25 & 10.40  & 10.88   &  -5   &   1   &  9    &   -9  &   9.68    \\
\hline
   106 &   165 & 271 &  0.22 &   &  8.83    &     &   3   &     &  -13  &   8.42    \\
   106 &   163 & 269 &  0.22 & 8.70  & 8.75 &      & -13 &     &   9   &   7.79    \\
   106 &   162 & 268 &  0.23 &   &  7.89    &      &     &     &       &           \\
   106 &   161 & 267 &  0.24 &   &  7.91    &      &   9   &     &   7   &   7.84    \\
   106 &   160 & 266 &  0.24 &   &  8.43    &      &       &     &       &           \\
   106 &   159 & 265 &  0.25 &   &  8.62    &      &   7   &     &  -11  &   8.51    \\
   106 &   158 & 264 &  0.25 &   &  8.92    &      &       &     &       &           \\
   106 &   157 & 263 &  0.25 & 9.40  &  9.02    &     &  -11  &     &   3   &   8.98    \\
   106 &   156 & 262 &  0.25 & 9.60  &  9.49    &     &       &       &       &           \\
   106 &   155 & 261 &  0.25 & 9.71  &  9.60    &     &   3   &     &   1  &   9.52    \\
   106 &   154 & 260 &  0.25 & 9.90  &  9.95   &       &       &       &       &           \\
   106 &   153 & 259 &  0.25 & 9.80  & 10.35    &     &   1   &     &  -9   &   9.57    \\
\hline
   105 &   158 & 263 &  0.25 &   &  8.26    &   9   &    &   -7  &     &   8.17    \\
   105 &   157 & 262 &  0.25 &   &  8.33    &   9   &  -11  &  -7   &   3   &   8.23    \\
   105 &   156 & 261 &  0.25 &   &  8.85    &   9   &    &  -7   &     &   8.73    \\
   105 &   155 & 260 &  0.25 &   &  8.94    &   9   &   3   &  -7   &   1   &   8.81    \\
   105 &   154 & 259 &  0.25 & 9.62  &  9.32    &   9  &    &  -7   &    &   9.18    \\
   105 &   153 & 258 &  0.25 & 9.50  &  9.72    &   9   &   1  &  -7  &  -9   &   8.75    \\
   105 &   152 & 257 &  0.25 & 9.21  &  9.03    &   9  &   &   -7  &     &   8.83    \\
   105 &   151 & 256 &  0.25 & 9.34  &  8.98    &   9  &  -9   &   -7  &   7   &   8.74    \\
\hline
   104 &   159 & 263 &  0.24 &   &  7.85    &    &   7   &    &  -11  &   7.74   \\
   104 &   157 & 261 &  0.25 & 8.65  &  8.37    &     &  -11  &     &   3   &   8.30    \\
   104 &   156 & 260 &  0.25 &        &  8.83    &     &       &     &       &           \\
   104 &   155 & 259 &  0.25 &   &  8.97    &     &   3   &     &   1  &   8.87    \\
   104 &   154 & 258 &  0.25 & 9.19  &  9.28    &       &       &       &       &           \\
   104 &   153 & 257 &  0.25 & 9.08  &  9.66    &    &   1   &     &  -9   &   8.75    \\
   104 &   152 & 256 &  0.25 & 8.93  &  8.93    &       &       &       &       &           \\
   104 &   151 & 255 &  0.25 & 9.06  &  8.90    &     &   -9  &     &   7   &   8.83    \\
\hline
   103 &   157 & 260 &  0.25 &   &  8.12    &   -7  &  -11  &   -1  &   3   &   7.60    \\
   103 &   156 & 259 &  0.25 &   &  8.57    &   -7  &      &   -1  &      &   8.10    \\
   103 &   155 & 258 &  0.25 & 8.90  &  8.78    &  -7   &  3    &  -1   &   1   &   8.12    \\
   103 &   154 & 257 &  0.25 &   &  9.04    &  -7   &      &  -1   &      &   8.50    \\
   103 &   153 & 256 &  0.25 &   &  9.45    &  -7   &   1   &  -1   &   -9  &   8.00    \\
   103 &   152 & 255 &  0.25 & 8.56  &  8.70    &  -7   &    &  -1   &      &   8.20    \\
   103 &   151 & 254 &  0.25 & 8.82  &  8.65    &  -7   &  -9   &  -1   &   7   &   8.07    \\
   103 &   150 & 253 &  0.25 & 8.92  &  9.10    &   -7  &    &  -1   &      &   8.65    \\
   103 &   149 & 252 &  0.25 & 9.16  &  9.25    &   -7  &   7   &  -1   &   5   &   8.62    \\
\hline
   102 &   155 & 257 &  0.26 & 8.48  &  8.08    &     &  3  &     &   1   &   7.97    \\
   102 &   154 & 256 &  0.25 & 8.58  &  8.36    &     &     &     &       &           \\
   102 &   153 & 255 &  0.26 & 8.43  &  8.72    &     &   1 &     &   -9  &   7.79    \\
   102 &   152 & 254 &  0.25 & 8.23  &  8.05    &     &     &     &     &           \\
   102 &   151 & 253 &  0.25 & 8.41  &  8.01    &     &  -9 &     &   7   &   7.90    \\
   102 &   150 & 252 &  0.25 & 8.55  &  8.53    &     &     &     &       &           \\
   102 &   149 & 251 &  0.25 & 8.75  &  8.73    &     &   7   &     &   5   &   8.48    \\
\hline
   101 &   157 & 258 &  0.26 & 7.27  &  7.09    &   -1  &  -11 &   7  &  3  &   6.80    \\
   101 &   156 & 257 &  0.26 & 7.56  &  7.49    &   -1  &      &   7  &     &   7.19    \\
   101 &   155 & 256 &  0.26 &        &  7.54    &   -1  &   3   &   7   &  1  &   7.13    \\
   101 &   154 & 255 &  0.26 & 7.91  &  7.79    &   -1  &       &   7   &     &   7.38    \\
   101 &   153 & 254 &  0.26 &        &  8.14    &   -1  &   1   &   7   &  -9 &   6.78    \\
   101 &   152 & 253 &  0.26 & 7.57  &  7.53    &   -1  &       &   7   &     &  6.99     \\
   101 &   151 & 252 &  0.26 &        &  7.50    &   -1  &  -9   &   7   &  7  &  6.82     \\
   101 &   150 & 251 &  0.25 & 7.96  &  8.05    &   -1  &       &   7   &     &   7.42    \\
   101 &   149 & 250 &  0.25 &        &  8.29    &   -1  &   7   &   7   &  5  &  7.37     \\
   101 &   148 & 249 &  0.25 & 8.44  &  8.54    &   -1  &       &   7   &     &     7.86  \\
   101 &   147 & 248 &  0.25 &        &  8.82    &   -1  &   5   &   7   &  1  &    7.87   \\
   101 &   146 & 247 &  0.25 & 8.76  &  8.81    &   -1  &       &   7   &     &    8.12   \\
   101 &   145 & 246 &  0.25 & 8.89  &  8.78    &   -1  &   1   &   7   & -7  &    8.04   \\

 \noalign{\smallskip} \noalign{\smallskip}

\end{tabular}

\end{ruledtabular}

\end{table*}


\endgroup



\begin{figure}
\includegraphics[width=\columnwidth]{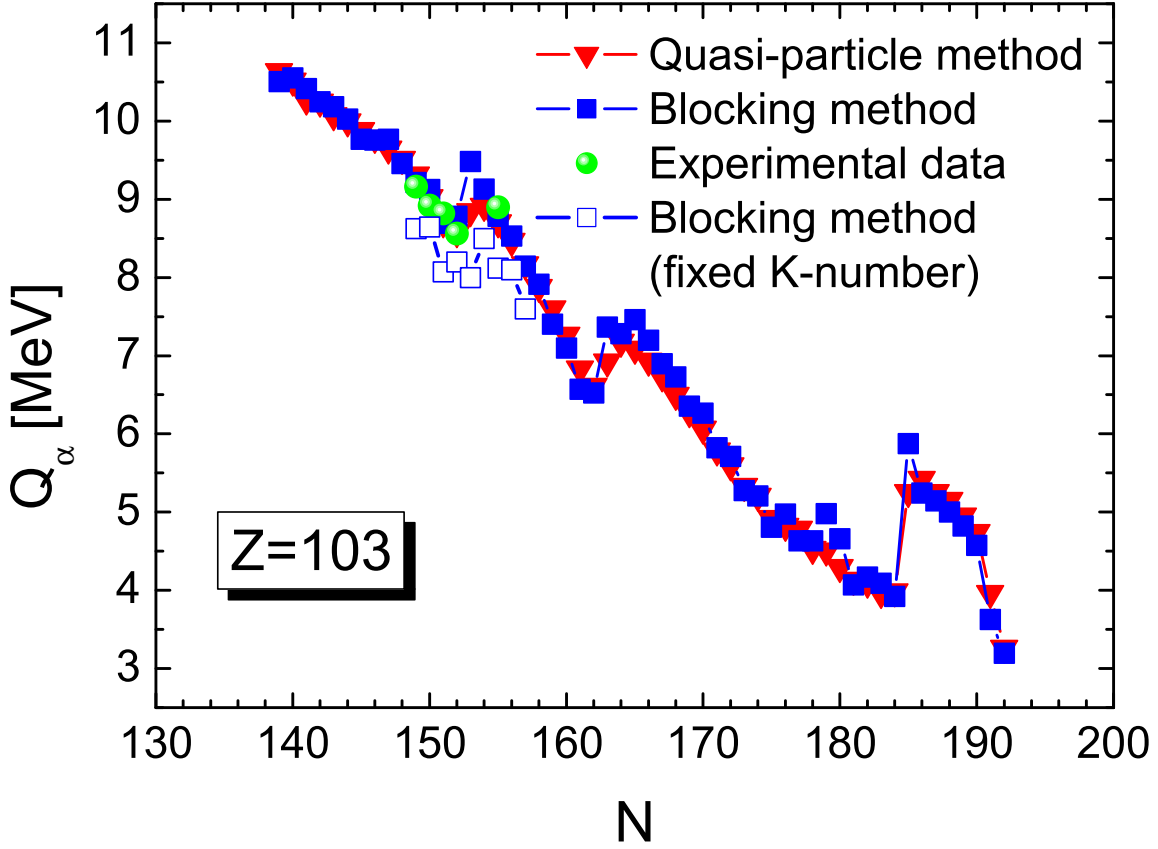}
\caption{\label{fig4}
  As in Fig. 1, but for Lr ($Z=103$) isotopes. Additionally, transition
  energies to the parent g.s. configuration in daughter,
 calculated with blocking, are shown as open squares. }
\end{figure}
\begin{figure}
\includegraphics[width=\columnwidth]{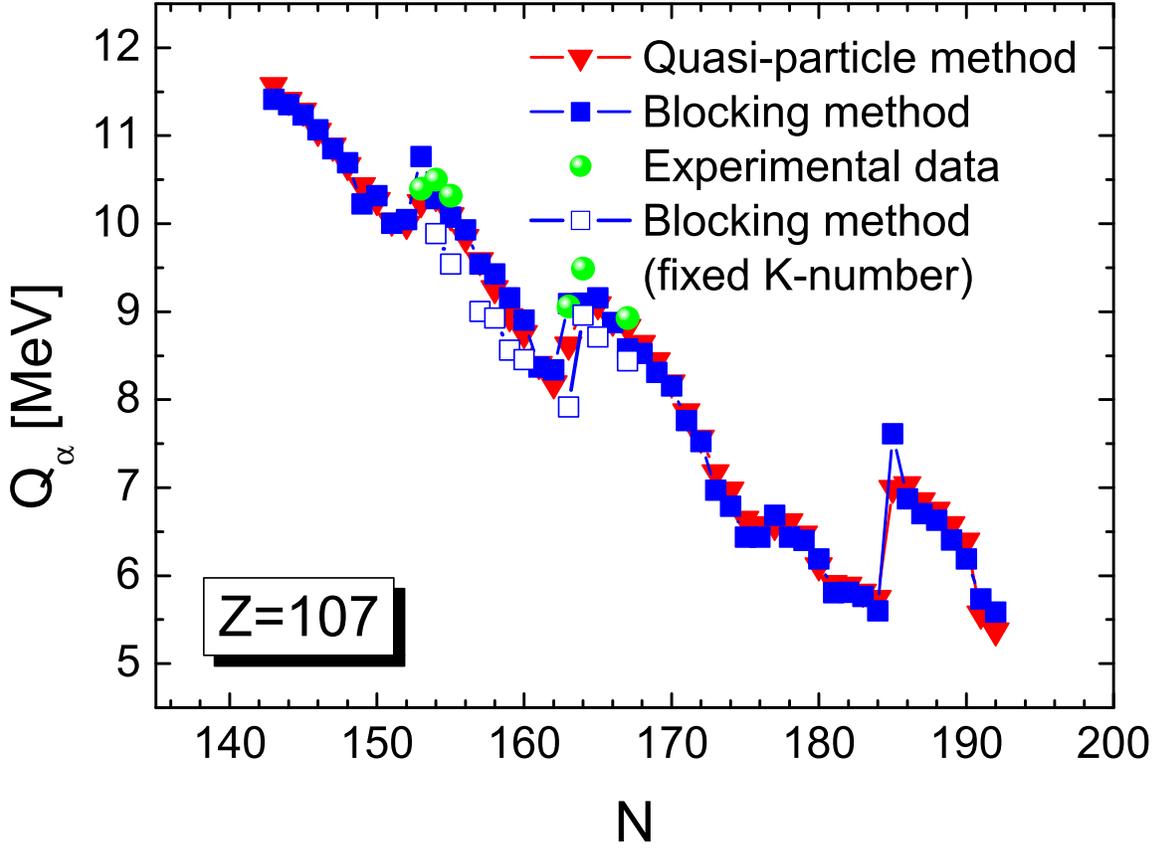}
\caption{\label{fig5} As in Fig. \ref{fig4}, but for Bh ($Z=107$) isotopes.}
\end{figure}
\begin{figure}
\includegraphics[width=\columnwidth]{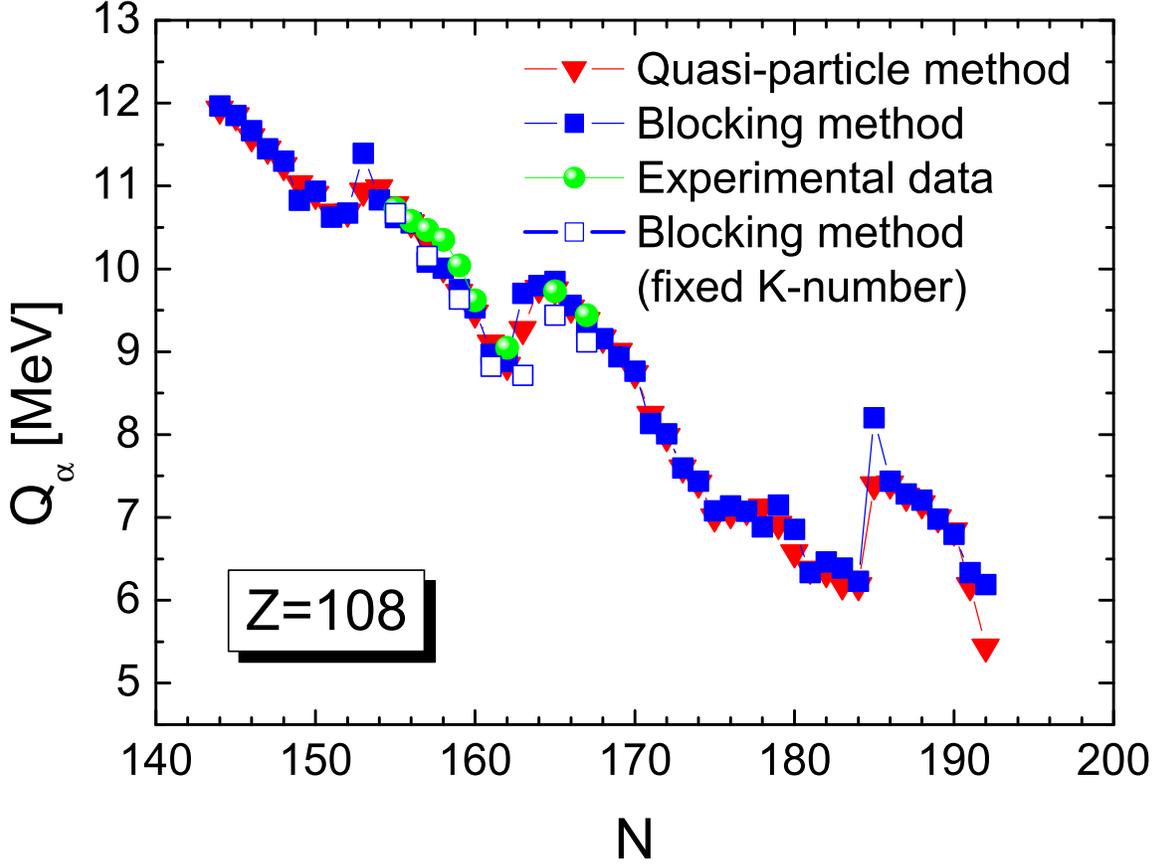}
\caption{\label{fig6} As in Fig. \ref{fig4}, but for Hs ($Z=108$) isotopes.}
\end{figure}
\begin{figure}
\includegraphics[width=\columnwidth]{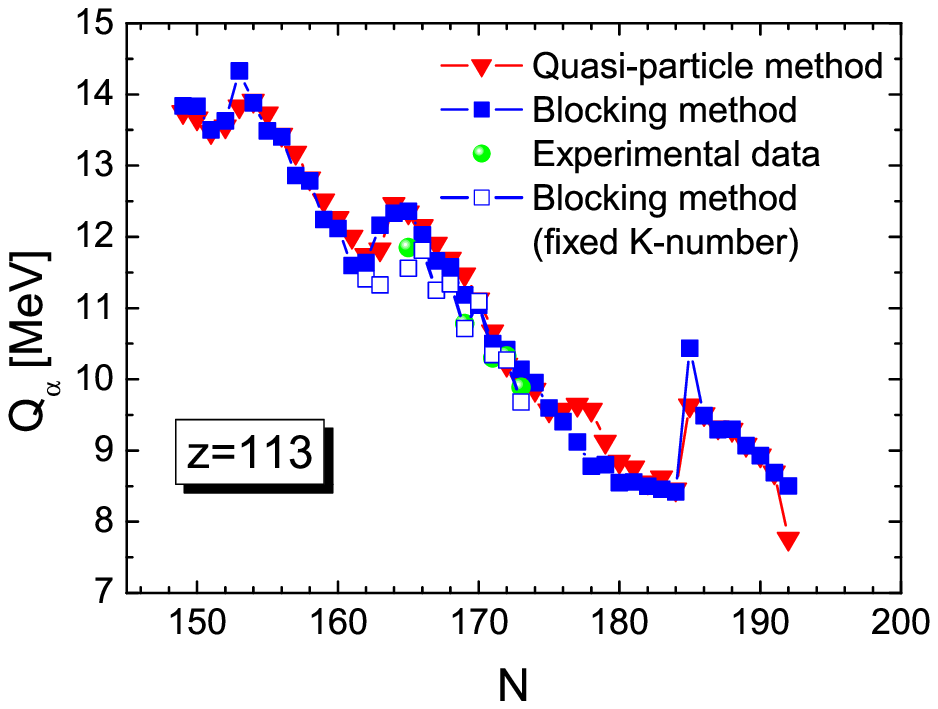}
\caption{\label{fig7} As in Fig. \ref{fig4}, but for $Z=113$ isotopes. }
\end{figure}


 The $Q_{\alpha}$ values are reasonably well reproduced for $Z=100-106$
 nuclei, where they can be larger or smaller than the experimental ones.
 Decay energies are slightly underestimated for $Z=108$ and systematically
 overestimated for $Z=111,112,113$.
 For nuclei with $Z\geq 101$, the mean deviation
  $<\mid Q_{\alpha}^{th}- Q_{\alpha}^{exp}\mid>$ and rms deviation
 are equal to, respectively, 217 keV and 274 keV in the calculation with
 blocking and 196 and 260 keV in the calculation with the quasiparticle
  method.  Thus, both methods give similar deviations which are smaller than in
  the region of the fit.
  Among 88 experimental $Q_{\alpha}$ values in Table V, 7 differ from the
 calculated ones by more than 0.5 MeV. In all cases the calculated values are
 too large. Five cases: $^{277,281}$Cn, $^{279,280}$Rg and $^{266}$Mt (the one
 with the largest deviation of 730 keV) signal the
  abovementioned overestimation of $Q_{\alpha}$ which
  somehow tends to disappear for the heaviest known parent nuclei (see Table V).
  A similar overestimate results from the calculation without
  blocking. For two other cases, the $N=153$ isotones of Rf and Sg (as well
  as of No, Db and Bh), the calculation without blocking gives results
  consistent with the experimental values. Thus, these two cases, as well
 as results for other $N=153$ isotones, should
  be understood as a specific failure of the calculation with blocking,
  described previously - the overshooting of the $Q_{\alpha}$ value just
 above the semi-magic gap $N=152$ (see Fig. 4-7).
   We have also checked the effect of a moderate 10\% increase in the pairing
  strengths on the $Q_{\alpha}$ values within the method of blocking in the
  seven cases mentioned above. It turns out that such a change
  mostly lowers $Q_{\alpha}$ values by less than 100 keV and increases
   one of them by nearly 200 keV.

\begin{figure}
\includegraphics[width=\columnwidth]{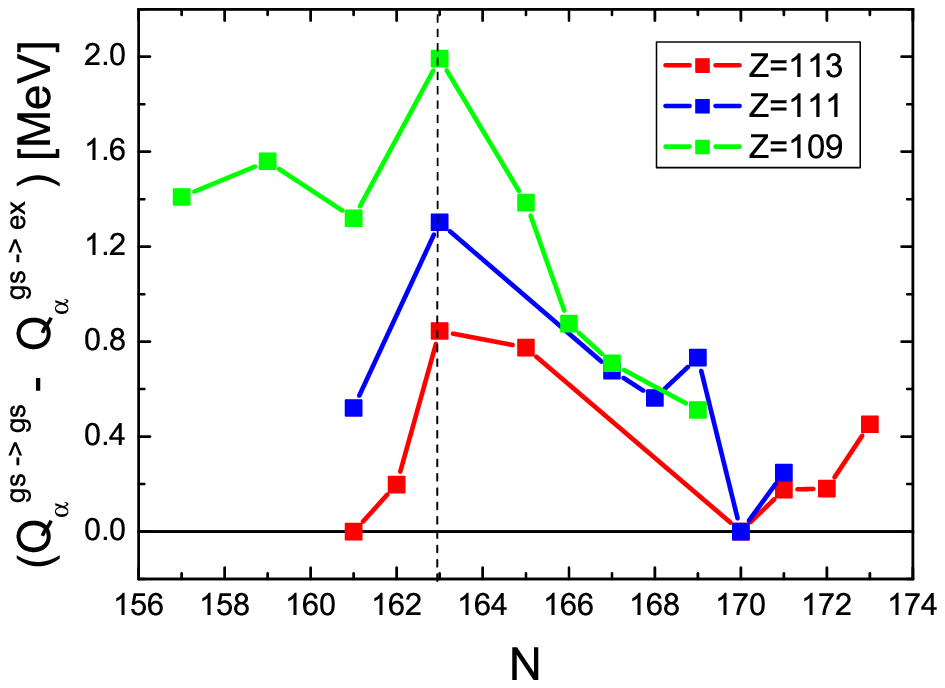}
\caption{\label{fig8} Excitation of the parent g.s. configuration
 in daughter nucleus calculated with blocking, equal to the predicted
  difference in transition energies, as a function of $N$
 for $Z=109$, 111, 113. }
\end{figure}

\begin{figure}
\includegraphics[width=\columnwidth]{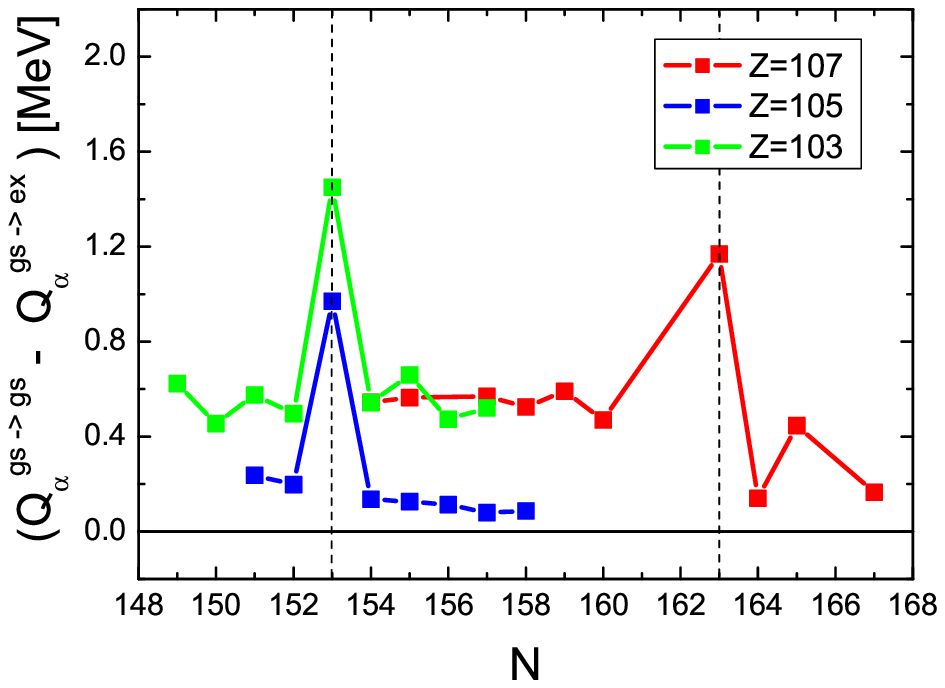}
\caption{\label{fig9} As in Fig. \ref{fig8}, but for $Z=103$, 105,
107. }
\end{figure}

  In trying to understand these results one has to remember that the g.s. to
  g.s. transitions are assumed in calculations. As mentioned in Sect. II,
  a predominance of transitions from the parent g.s. to an excited state in
  the daughter nucleus may result in attributing an apparent $Q_{\alpha}$ value
  lower than the true one. If one assumes that the $\alpha$ decay proceeds to
  the parent g.s. configuration in the daughter, one obtains energies shown in
  the last column in Table V.
  It may be seen that energies of these configuration-preserving transitions
  are reduced especially for particle numbers corresponding to one particle
  above a closed subshell. Predicted energies of configuration-preserving
  transitions are also shown in Fig. 4-7 for four isotope chains.
  A particularly large, about 2 MeV excitation of the parent configuration
  occurs in the daughter of $^{272}$Mt; the excitation of
  about 1.5 MeV occurs for $Z=103$, $N=153$. These excitations, equal to
  the differences between g.s.$\rightarrow$g.s. and configuration-preserving
  transition energies, $Q_{\alpha}^{gs -> gs}-Q_{\alpha}^
  {gs -> ex}$, are shown in Fig. \ref{fig8},\ref{fig9}.
  The particle numbers $Z=103,109$ and $N=153,163$ correspond to s.p.
  orbitals lying just above the large gaps.

  This is illustrated in Fig. \ref{fig10}, where neutron and proton s.p.
 energies are shown vs $\beta_2$. The deformation $\beta_4$ was chosen to
  roughly follow g.s. minima around $Z=109$, $N=163$ ($\beta_2=$0.22,
  $\beta_4$=-0.08): oblate minima for $Z\geq 115$ correspond to small
  $\beta_4$, prolate minima for $Z=106,107$ have $\beta_4\approx -0.05$.
  As other deformations, differences in $\beta_4$ between isotopes and
  $Z$ and $N$-dependence are omitted, Fig. \ref{fig10} can serve only a general
  orientation. It may be seen, that above $N=162$ and $Z=108$ the Woods-Saxon
  model predicts  two intruder orbitals: neutron $K^{\pi}=13/2^-$ and proton
  $11/2^+$. Similarly, the intruder neutron $K^{\pi}=11/2^-$ and proton
  $9/2^+$ orbitals lie above $N=152$, $Z=102$. In general, such orbitals could
  combine spins and form a high-K isomer; for $Z=109$, $N=163$, our model
  predicts such a configuration as a ground state. A substantial hindrance of
  the g.s. to g.s. $\alpha$-decay could be expected in such case. Then,
  it is also not excluded that the g.s. decay would be so hindered, that
   the $\alpha$ decay would proceed from an excited state.
   Only future experimental data may show whether
  considering such a possibility will be necessary.

   The predicted neutron and proton g.s. configurations are given in Table V,
   both for parent and daughter nuclei. They can be compared to the measured
  ones only in a few cases. For example, the $3/2^+$ g.s. of $^{257}$No
  \cite{Asai} is reproduced in our calculation. On the other hand, the
  predicted ground states in $^{255}$Lr and $^{101}$Md are interchanged with
   respect to the experimental results \cite{Chat}.
   Ground state spins and parities evaluated from measurements in other
   Md, No, Lr and Rf isotopes are consistent with our calculations, except
   for the measured or evaluated $7/2^-$ ground states in Md.
 The proton configurations predicted by the quasiparticle method are the
   same as in Table V. Mostly it is also the case for neutrons, except for
   the 155-th neutron being $1/2^+$ instead of $3/2^+$. The g.s.
  configurations in odd-A actinides, calculated within
  the Woods-Saxon model with the quasiparticle method, may be found in \cite{ParSob}.

\begin{figure}[!h]
\begin{minipage}{60mm}
\includegraphics[scale=0.3]{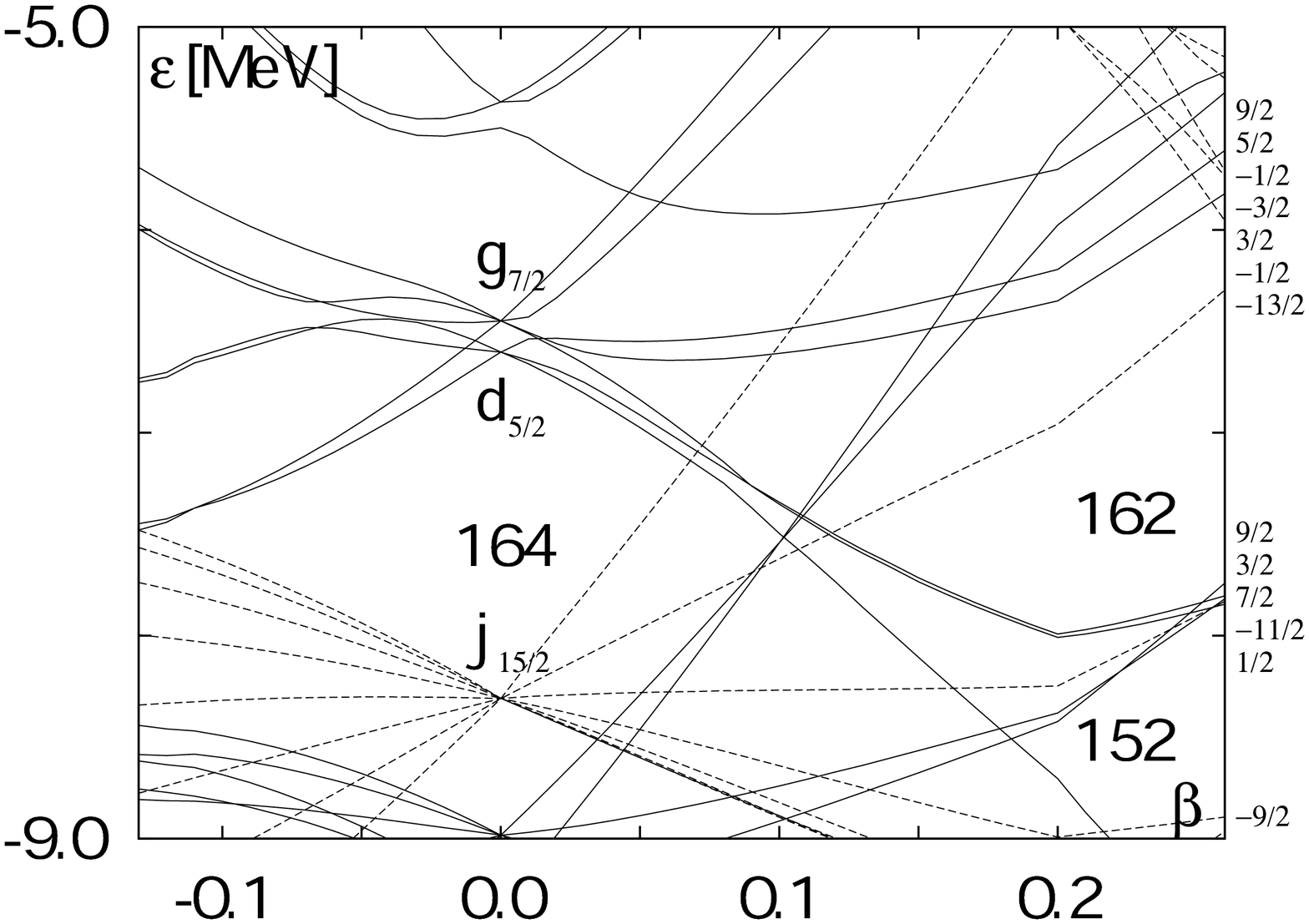}
\end{minipage}
\begin{minipage}{60mm}
\includegraphics[scale=0.3]{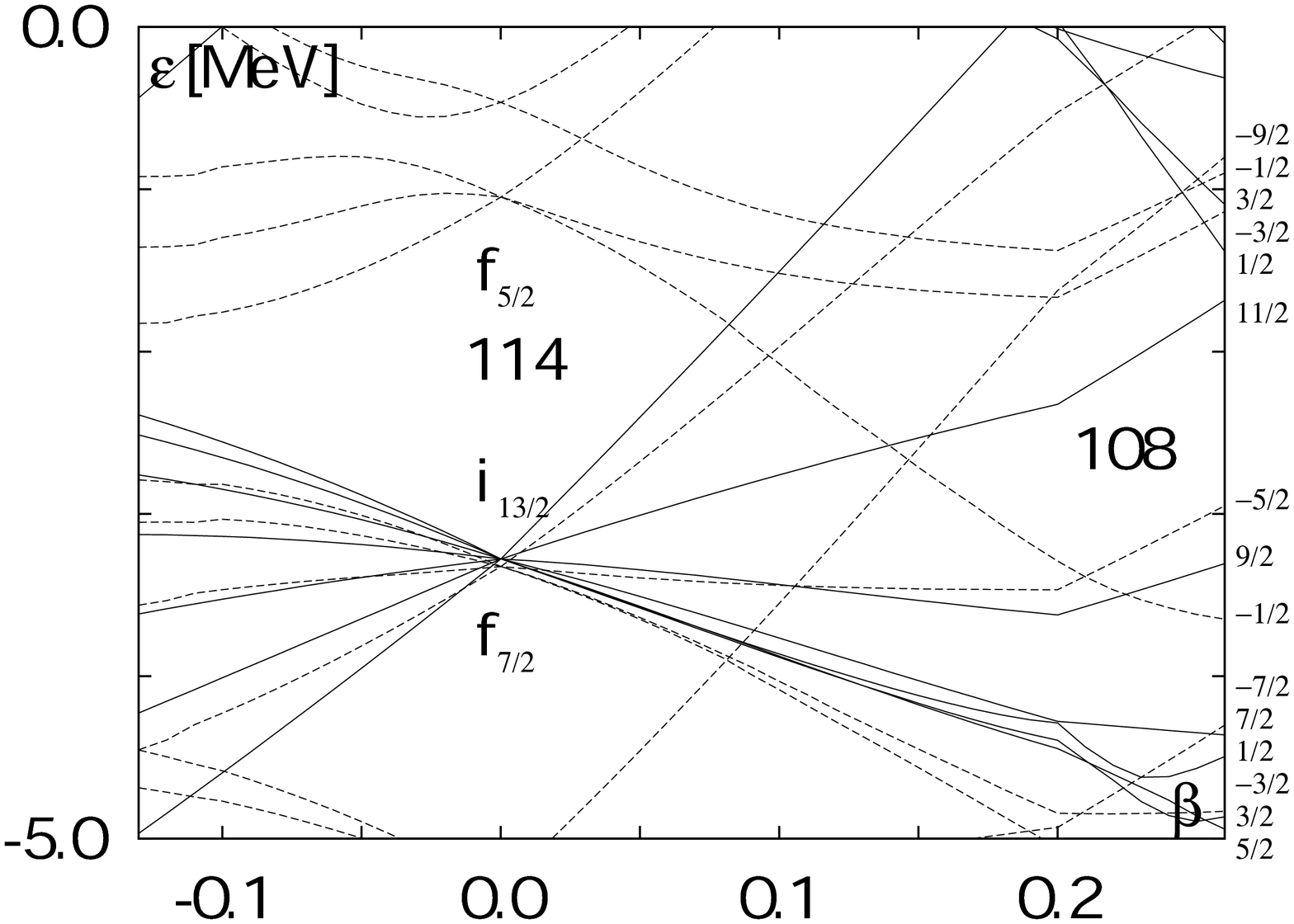}
\end{minipage}
\caption{\label{fig10} Single-particle levels for neutrons (top) and protons
  (bottom) for $Z=109$ $N=163$. To roughly follow ground-state minima,
  deformation $\beta_4=0$ for $\beta_2<0$, then decreases linearly down
  to -0.08 at $\beta_2=0.2$ and then linearly rises to
  -0.05 at $\beta_2=0.26$. Positive (negative) parity levels are drawn with
  solid (dashed) lines; quantum numbers $\pi, K$ are given to the right. }
\end{figure}

\begin{figure}
\includegraphics[width=\columnwidth]{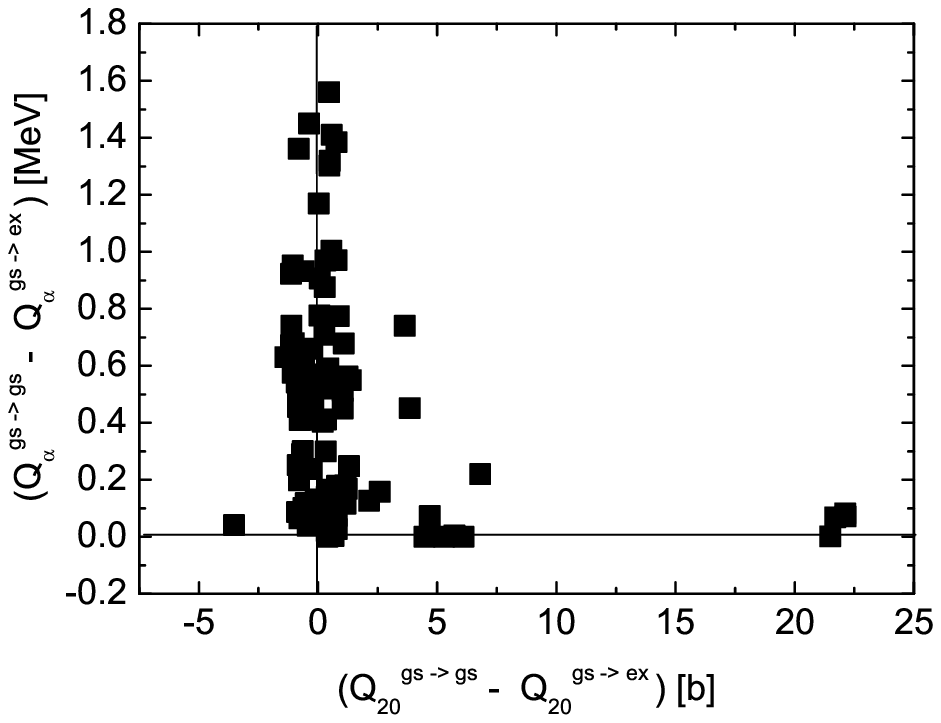}
\caption{\label{fig11} Correlation between calculated parent-daughter
   deformation difference and the excitation of the parent g.s. configuration
  in daughter nucleus. }
\end{figure}

  The g.s. to excited state transitions could also result from a deformation
  difference between parent and daughter. Such changes happen for some
  $Z\geq 114$ parent nuclei (weakly oblate to weakly prolate) and for
  parents with $Z=111-113$, $N\approx 169$ (increase in prolate deformation).
  As can be seen in Fig. \ref{fig11}, the correlation between calculated
   deformation change and the excitation of the parent configuration in
  daughter is weak for nuclei investigated here:
 a large difference in quadrupole moments of the parent and
  daughter is not accompanied by a large change in the transition energy.

\begin{figure}
\includegraphics[width=\columnwidth]{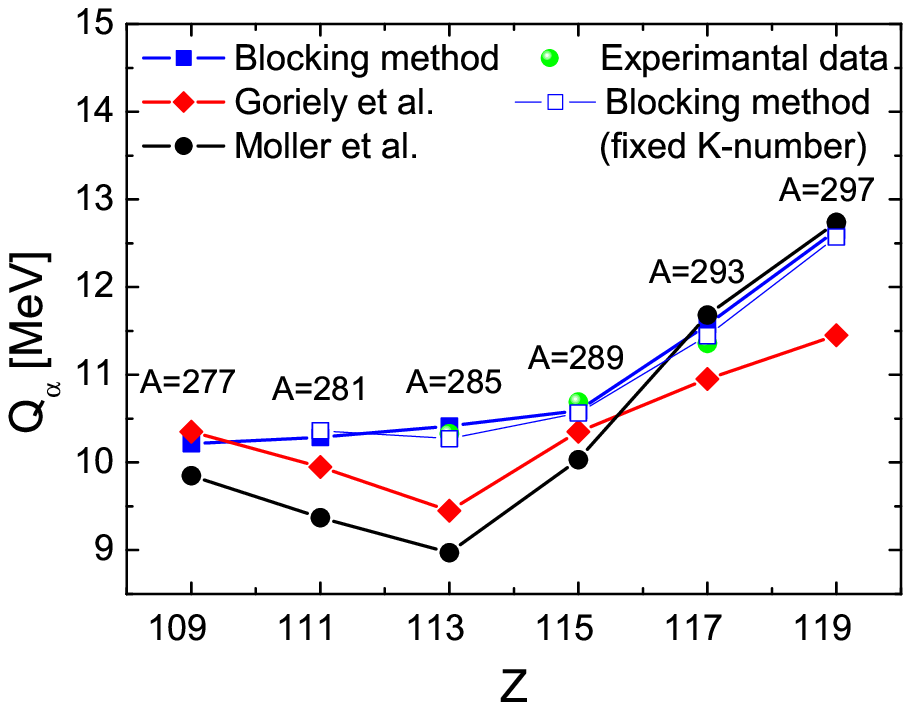}
\caption{\label{fig12} $Q_{\alpha}$ values in a chain beginning
 at a hypothetical nucleus $Z=119$, $A=297$, which contains the known chain
  for $^{273}$117 - models vs experiment. }
\end{figure}

  The results of the model can also be appreciated by comparing successive
  transitions along the measured alpha decay chains.
  For the recently measured $^{294}$117 and $^{294}$118 chains
 \cite{Oga2012} this is done in Figs. 12-15. The partially known decay chain
  for a hypothetical nucleus $^{297}$119 is shown in Fig. 11.
 The data were taken
 from \cite{Gupta2005,Oga2012} and some other sources \cite{O2,Rudolph}.

 After the successful synthesis of elements Z=117 and Z=118,
 the hypothetical nucleus $^{297}$119 is a
 natural candidate for the next synthesis experiment.
 One of the likely reactions leading to this element seems to be
 $^{48}$Ca($^{252}$Es,3n)$^{300-x}$119 see eg. \cite{Wilczynska}. Note, that
 this $\alpha$-particle chain contains the known decay chain of
 $^{293}$117 \cite{Oga2012}. One can see that our results
 reasonably agree with the experimental data and HFB-14
 predictions \cite{Gortable}. However, for the nucleus $^{297}119$
 our result is close to the model of Moller et al. \cite{mol},
 which underestimates $Q_{\alpha}$ values for lighter nuclides in the
 chain.
\begin{figure}
\includegraphics[width=\columnwidth]{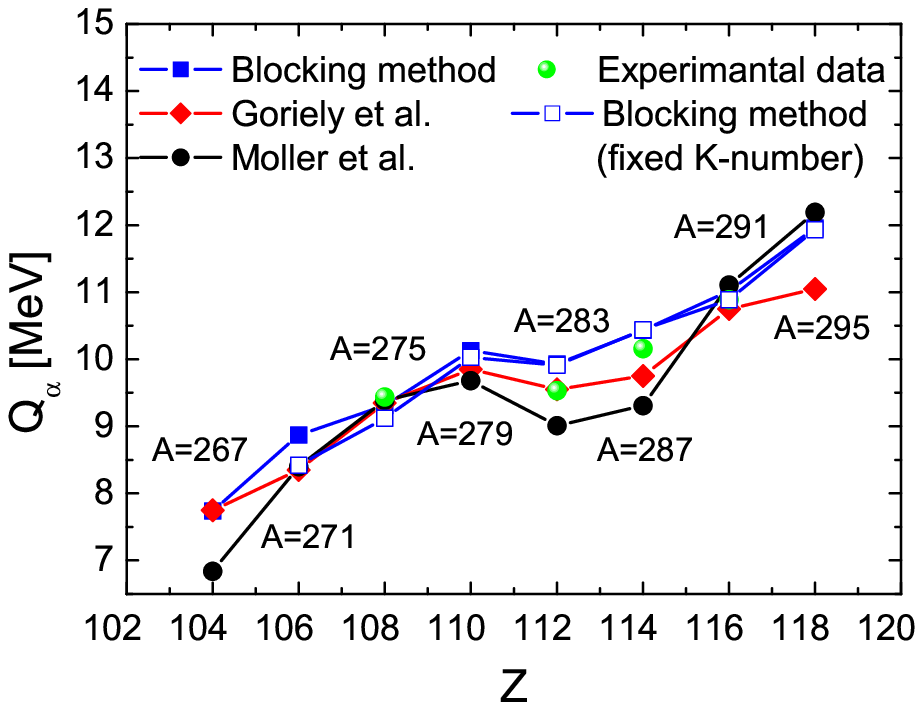}
\caption{\label{fig13} As in Fig. \ref{fig12}, but for $Z=118$, $A=295$.}
\end{figure}

 Another comparison is given in Fig. \ref{fig13} for the $\alpha$-particle
 chain starting at $^{295}$118. It may be seen that, compared to \cite{mol},
 a similar or better (especially for $^{283}$112 and
 $^{287}$114) agreement with the data is obtained by the present
 model. A similar conclusion follows when comparing the present
 results to the self-consistent calculations \cite{Gortable}.
 Note, that this $\alpha$-particle chain contains
 the well-known chain of element $^{291}$116.
\begin{figure}
\includegraphics[width=\columnwidth]{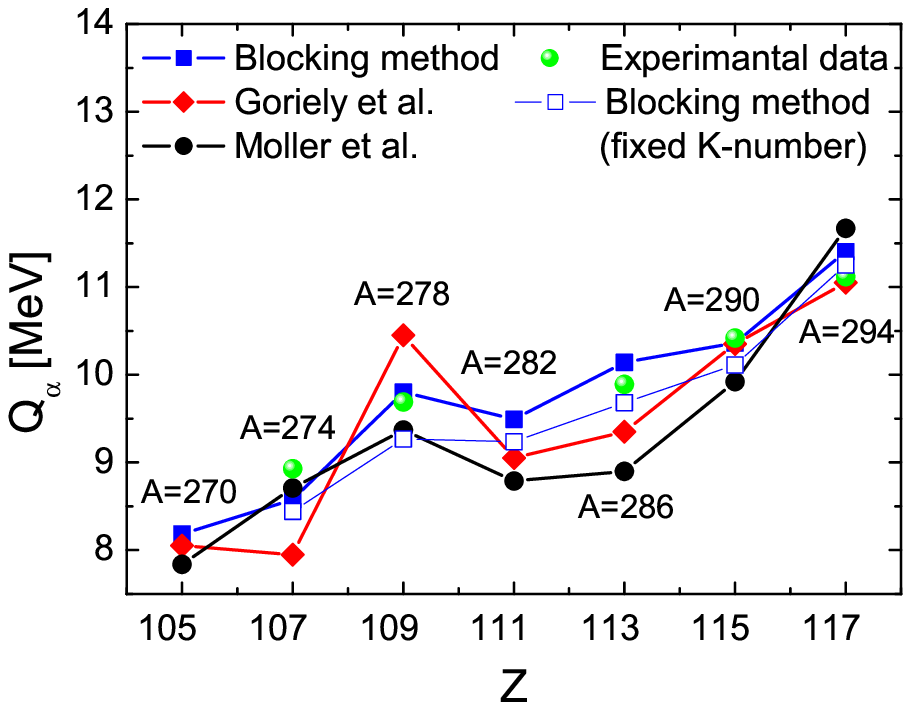}
\caption{\label{fig14} As in Fig. \ref{fig12}, but for $Z=117$, $A=294$.  }
\end{figure}
\begin{figure}
\includegraphics[width=\columnwidth]{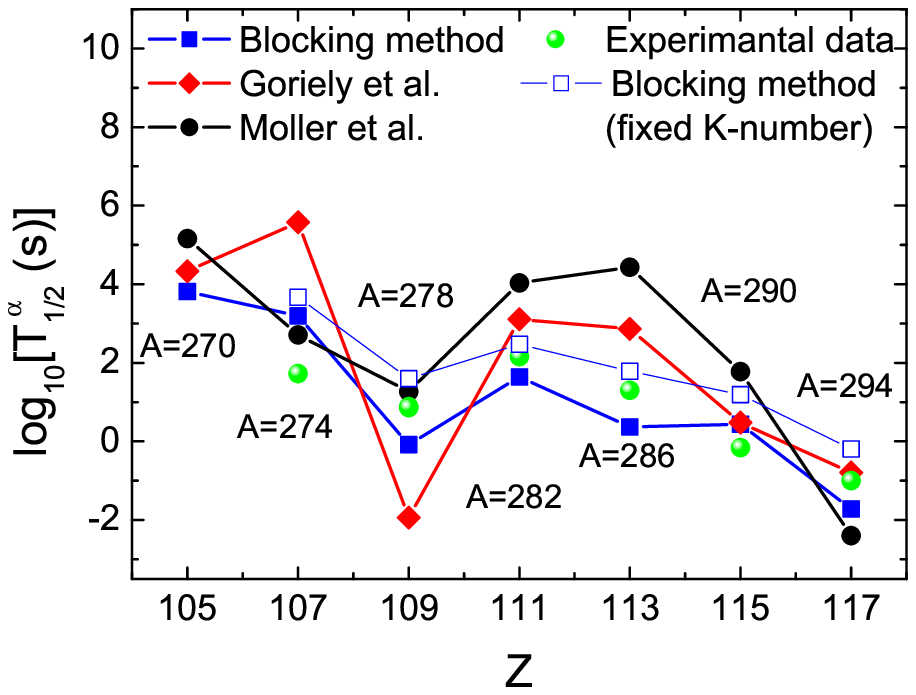}
\caption{\label{fig15} Calculated vs experimental alpha half-lives for
  the decay chain of $Z=117$, $A=294$. }
\end{figure}

 As an example of odd-odd systems, the alpha-chain for the nucleus
 $^{294}$117 is shown in Fig \ref{fig14}. One may notice a good agreement
 between our $Q_{\alpha}$-values and the recently reported experimental data
 \cite{O2,Oga2012}. The other models deviate more from the measured
 $Q_{\alpha}$-values in this chain. This has an impact on
 the predicted alpha-decay lifetimes, as shown in Fig. \ref{fig15}.
 For example, in case of the HFB-14 approach \cite{Gortable},
 the half life of $^{274}$Bh is overestimated by four orders of magnitude
 while the one for $^{278}$Mt is underestimated by three
 orders of magnitude. The half-lives resulting from \cite{mol} are
 systematically overestimated as a consequence of the underestimated
 $Q_{\alpha}$-values. (The Viola-Seaborg-type formula from
  \cite{Royer} has been used to convert $Q_{\alpha}$ to half-lives).
 In all three discussed chains, our results are slightly overestimated.
 At present, however, the explanation that the allowed decays go to the excited
 states (lying slightly above the ground state), is not excluded,
 especially in the context of recent spectroscopic studies of element
 $Z=115$ by Rudolph et. al \cite{Rudolph}.



 \section{Conclusions}

 A systematic calculation of nuclear masses in the region of superheavy
 nuclei, including odd and odd-odd systems, was performed within the
  microscopic-macroscopic model with the Woods-Saxon deformed potential.
 Two versions of the model were used, with and without blocking.
 A fit in the region of heavy nuclei was performed to fix 3 additional
 parameters of the model, one for each group of odd-A and odd-odd
  nuclei, while keeping all previous parameters as they were used
  for even-even nuclei. Then, the $Q_{\alpha}$ values were calculated for
 SH nuclei as a prediction of the model to be compared against the data.
 The quality of the prediction turns out better than the quality of the
 model in the region of the fit: in the version with
  blocking, the mean and rms deviations of 217 and 274 keV for 88 SH nuclei
  are smaller than 326 and 426 keV for the 204 nuclei from the fit region.
  The quasiparticle method, clearly better in the region of the fit,
  for SH nuclei gives similar mean and rms deviations of 196 and 260 keV
  as the calcutions with blocking.

  Both versions of the model similarly overestimate $Q_{\alpha}$ values
  for $Z=111-113$ and underestimate them, although to a lesser extent,
  around $Z=107,108$. At present, these result should be treated with some
  care. Many of synthesized SH isotopes are odd-$A$ or odd-odd nuclei and
  in many cases the statistics of $E_{\alpha}$ values is not large.
  Therefore it is not completely clear whether some
  of those cases may be explained by a hindrance of g.s. to g.s. transitions.
  The g.s. configurations of some SH nuclei, especially those involving high-K
  intruder orbitals, strongly hint to a possibility of $\alpha$-decay
  hindrance, for example, for $Z=109$, $N=163$.

  As comparisons to other models show, the agreement with data obtained
  here, without any parameter adjustment for $Q_{\alpha}$ and with a minimal
  adjustment for masses, is surely not worse.
 This gives a confidence that some refinements, especially in the
  treatment of pairing, may still moderately improve the agreement with data.





\begin{thebibliography}{99}
\bibitem{Ealpha}
 The alpha particle energy is in direct relationship with Q-value
 $E_{\alpha}=\frac{A-4}{A}Q_{\alpha}$ where A is the mass number.
\end{thebibliography}
\end{document}